\documentclass[aps,prb,reprint,superscriptaddress]{revtex4-2}

\usepackage{hyperref}
\usepackage{xcolor}
\usepackage{graphicx}
\usepackage{subcaption}
\usepackage{amsmath}
\usepackage{amsfonts}
\usepackage{physics}
\usepackage{dsfont}

\graphicspath{{./figures/}}

\begin{document}

	\title{Disordered Graphene Ribbons as  Topological Multicritical Systems}

	\date{\today}

	\author{Saumitran Kasturirangan}
	\affiliation{School of Physics and Astronomy, University of Minnesota, Minneapolis, Minnesota 55455, USA}
	\author{Alex Kamenev}
	\affiliation{School of Physics and Astronomy, University of Minnesota, Minneapolis, Minnesota 55455, USA}
	\affiliation{William I. Fine Theoretical Physics Institute, University of Minnesota, Minneapolis, Minnesota 55455, USA}
	\author{Fiona J. Burnell}
	\affiliation{School of Physics and Astronomy, University of Minnesota, Minneapolis, Minnesota 55455, USA}

	\begin{abstract}
		The low energy spectrum of a zigzag graphene ribbon contains two gapless bands with highly non-linear dispersion, $\epsilon(k)=\pm |\pi-k|^W$, where $W$ is the width of the ribbon. The corresponding states are  located at the two opposite zigzag edges. Their presence reflects the fact that the clean ribbon is a quasi one dimensional system naturally fine-tuned to the topological {\em multicritical} point. This quantum critical point separates a topologically trivial phase from the topological one with the index $W$. Here we investigate the influence of the (chiral) symmetry-preserving disorder on such a multicritical point.  We show that the system harbors delocalized states with the localization length diverging at zero energy in a manner consistent with the $W=1$ critical point. The same is true regarding the density of states (DOS), which exhibits the universal Dyson singularity, despite the clean DOS being substantially dependent on $W$. On the other hand, the zero-energy localization length critical exponent, associated with the lattice staggering, is not universal and depends on the topological index $W$.     	
	\end{abstract}

	\maketitle

	\section{Introduction}

	Disorder is generically relevant in low-dimensional systems, leading to localization and a lack of transport even for arbitrarily weak disorder.  In two dimensions, disorder is marginally relevant, 
and perturbations such as the addition of spin-orbit coupling allow for the possibility of a metal-insulator transition at a finite disorder strength \cite{AndersonLocalization,LeeRamakrishnanReview,KramerMackinnonReview,EversMirlinReview}. In one dimension, however, the scaling theory of localization predicts that disorder is relevant\cite{ScalingTheoryLoc,GorkovLarkinScaling}, and localization is always expected in the presence of disorder.

	If the conditions are just right, it is possible to observe a critically delocalized state in quasi-1D, similar to the state at a 2D metal-insulator transition. This state is present precisely at the transition between topologically distinct phases. The simplest example of this phenomenon is the 1D chain with random nearest neighbor hopping, first studied by Dyson \cite{DysonDOS}. In the absence of disorder, if the nearest neighbor hoppings are staggered uniformly, say due to Peierls instability, the system is in an insulating phase with a winding number (topological invariant) either $0$ or $1$ \cite{SSH,ZakPhase}.  When all the hoppings are identical, there is a linearly dispersing gapless mode near zero energy and the system is at the critical point between two topologically distinct phases. This critical point survives in the presence of hopping disorder, provided that all bonds are identical on average: even in the presence of disorder,  there must be a critically delocalized state at zero energy at the boundary between phases with different average band structure topology.  

The mechanism for the formation of the delocalized state is deeply rooted in topology \cite{MotrunichDamleHuse}. Adding disorder creates domains between the topologically distinct phases, with low-energy bound states localized on these domain walls. The interactions of the low-energy bound states is what gives rise to the critical state at zero energy, which is known to have multifractal characteristics \cite{BalentsFisher,Prodan2014,Kamenev2016}. 
  This mechanism also leads to critical zero-energy states in 1D Dirac Hamiltonians with random mass, and in random XY spin chains \cite{BalentsFisher,ShankarXY,McKenzieQPT}.

While the localization length of the state at zero energy is formally infinite, it is not a perfectly conducting delocalized state, but one with a broad transport distribution that is dominated by rare-region effects.   The associated transport statistics at zero energy have traditionally been  studied using the Fokker-Planck approach, which describes the evolution of the probability distributions of transport with the length \cite{Dorokhov82,MPK,MacedoChalker,BeenakkerRMT}.  For Hamiltonians in symmetry class BDI (AIII), which describes the 1D chain with random nearest neighbor hopping when  time-reversal symmetry is present (absent), 
 this approach predicts that  at zero energy, in 1D systems with   an odd number of channels, the average conductance does not decay exponentially, but instead falls off as a power of the system size \cite{Deloc1DChains,MudryRandomFlux}.  

In addition to BDI and AIII, the 10-fold classification scheme for topological insulators and superconductors \cite{AltlandZirnbauer,RyuTenFoldWay,KitaevClassification,LudwigClassification} identifies several other symmetry classes that admit multiple topologically distinct phases in 1D.   These classes admit gapped phases with an integer (BDI, AIII, and CII) or $\mathbb{Z}_2$ (D and DIII) valued topological invariant in 1D, and thus also contain Hamiltonians describing the critical point separating these. 
All of them correspondingly harbor a disordered critical point at zero energy, whose transport can be studied using the Fokker-Plank approach \cite{BrouwerFPReview}.  This predicts a power-law in system size average conductance at zero energy in classes BDI, AIII, and CII for an odd number of channels, and in classes D and DIII for any number of channels \cite{LocalizationDelocalizationDirtySC}. 

In reality,  for all of these symmetry classes accessing the true critical point requires fine-tuning \cite{MotrunichDamleHuse,GruzbergSuperuniversality}, and   
 in general, it is possible to add symmetry allowed terms not present in the Fokker-Plank treatment that move the system away from criticality.  Interestingly, however, sufficiently close to the critical point the transport statistics are found to be universal across all symmetry classes in 1D \cite{GruzbergSuperuniversality}.



The present work focuses on zigzag graphene ribbons with nearest neighbor hopping. 
Although these belong to the same symmetry class (BDI) as armchair ribbons and the random hopping chain, in the clean limit they have very different properties near zero energy \cite{BernevigHughesGrapheneChapter,WakabayashiGrapheneRibbons}: 
they have a higher-order band crossing at zero energy, with a dispersion of the form $\abs{\pi-k}^W$ for momenta $k$ close to $\pi$ (the band crossing point), where $W$ is the width of the ribbon. Therefore the lowest energy sub-band gets increasingly flat as the width, $W$, increases. This reflects the fact (see section \ref{subsec:topology}) that a clean zigzag ribbon  is at a multicritical point, i.e. it is at the transition point between topological phases whose winding numbers differs by $W$. Our goal is to study the fate of this multi-critical point in the presence of hopping disorder.    

Because of the higher-order band crossing, transport strictly at zero energy is not accessible, since the velocity of 0-energy particles incident from a clean lead is $0$.  Instead, we focus on the system's behavior as  the energy $\epsilon \rightarrow 0$.   For typical disordered critical points, this exhibits many features indicative of the nature of the underlying criticality.   The density of states has a characteristic divergence, of the form $1/(\abs{\epsilon \ln^3(\epsilon)})$  \cite{DysonDOS,TitovDOS,BrouwerFPReview}. The {\it typical} localization length diverges logarithmically  with energy \cite{Theodorou1DChain, EggarterLocDivergence, ZimanLocDivergence, DFisherIsingChains, McKenzieQPT,Kamenev2014}. Rare-region effects play an increasing role closer to the critical point \cite{Kamenev2016}, and the {\it average} properties start to differ dramatically from the {\it typical} ones. For instance, the {\it average} localization length diverges as $\ln^2(\epsilon)$ \cite{BalentsFisher}.  We will show that these divergences also describe the low-energy regime of the disordered multi-critical point relevant to zig-zag graphene ribbons.  

The traditional Fokker-Planck approach cannot describe these divergences, as away from zero energy it ceases to capture the transport statistics of disordered critical points. 
 To access this regime, the 1-parameter scaling inherent to the Fokker-Plank equation must be replaced by a 2-parameter scaling, 
 describing the crossover of transport statistics from the critical point at zero energy (described by the Fokker-Plank equation for class BDI) to that at high energies (described by a Fokker-Plank equation for class AI)\cite{RyuCrossover,TwoParamScaling}.  
 This scaling leads to a universal form of the transport distribution, which captures the low-energy regime in both the 1D chain with hopping disorder, and in metallic arm-chair graphene ribbons \cite{TwoParamScaling},
 which in the clean limit also have a linearly dispersing band crossing at zero energy associated with a critical point separating topologically distinct phases.   Here, we give evidence that the same 2-parameter scaling, and underlying distribution, describe transport at low energies in the zig-zag case.

An alternative perspective on the higher-order dispersion in zig-zag graphene ribbons stems from the fact that two-dimensional graphene is an example of a topological semi-metal: the large number of very low energy states of the zigzag graphene nanoribbon, which are localized near its edges, are a manifestation of the boundary flat band in the 2D topological semi-metal. 
The fate of the boundary modes in the presence of disorder is also investigated. These are found to be stable and remain close to zero energy when hopping disorder is added, indicating that this feature of the topological semi-metal is robust to disorder.


	There are several works which have studied the effects on disorder on zigzag graphene ribbons, though not for the current scenario, where hopping disorder is considered. A few of these works are highlighted here to provide a broader perspective. In \cite{WakabayashiPerfectlyConductingChannel, WakabayashiPerfectlyConductingChannel2,MuccioloPerfectlyConductingChannel}, it was found that there is a perfectly conducting channel in zigzag ribbons in the presence of {\it long-range on-site} disorder. This is because there is an additional chiral mode in the each valley that is not affected by long-range impurities. However, for short-range impurities, localization is still expected, as was shown in \cite{LocalizationinGrapheneLengthScales,MetalInsulatorTransitionGNR,LocalizationandBKTGraphene}. Disorder on the edge has also been found to strongly affect the transport in graphene ribbons \cite{MuccioloEdgeDisorder,KatsnelsonEdgeDisorder}. In \cite{GrapheneQuantumDotDisorder}, it was shown that the edge states are stable under the presence of edge roughness.

	The rest of this paper is structured as follows. In section \ref{sec:background}, the model of zig-zag graphene with only nearest neighbor hopping is introduced, followed by the properties of the spectrum in the absence of disorder. Particular emphasis is placed on the low-energy band and its dispersion. This is followed by a discussion of the generic symmetries of the model and its topological properties. In section \ref{sec:disorder}, the transport, density of states, and stability of the edge states of the disordered zigzag ribbon are discussed. This is followed by a discussion of the results in section \ref{sec:conclusions}.

	\section{Spectrum, symmetry, and topology in zigzag graphene}\label{sec:background}

%

	\subsection{Model and spectrum}\label{subsec:model-and-spectrum}

\begin{figure}
		\centering
		\includegraphics[width=\linewidth]{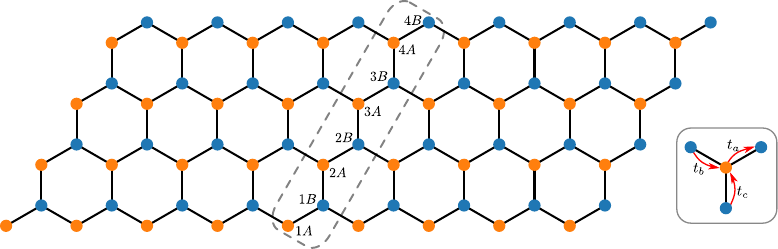}
		\caption{Zigzag graphene ribbon of width $W=4$ is shown. Differently colored sites correspond to different sub-lattices and the dashed box encloses the unit-cell of the zigzag ribbon which is translation invariant in the horizontal direction. The labeling convention for the sites used in this work is also shown here. The inset shows the hopping parameters.}
		\label{fig:zz-lattice}
	\end{figure}

	The present work focuses on  zigzag graphene ribbons 
	with nearest-neighbor hopping, described by the Hamiltonian:
	\begin{multline}\label{eqn:ham-zz-real-space}
			H = \sum_i \sum_{j=1}^{W} t_{i, j}^{a} c^{\dagger}_{i,j,B} c_{i, j, A} + \text{h.c.} \\
			+ \sum_i \sum_{j=1}^{W} t_{i, j}^{b} c^{\dagger}_{i+1,j,A} c_{i, j, B} + \text{h.c.} \\
			+ \sum_{i} \sum_{j=1}^{W-1} t_{i, j}^{c} c^{\dagger}_{i,j+1,A} c_{i, j, B} + \text{h.c.} 
		\end{multline}
	where the hopping parameters $t_{i, j}^{a,b,c}$ are real. The choice of unit cell, and conventions for labeling sites, are shown in Figure \ref{fig:zz-lattice}. Here, $c^\dagger_{i, j, \alpha}$ is the creation operator for an electron on the unit cell labeled by index $i$ along the horizontal direction, on the vertical chain labeled by $j$; the subscript $\alpha \in \{A, B\}$ refers to the $A$ and $B$ sub-lattices of the 2-site unit cell of the honeycomb lattice (orange and blue, respectively, in Figure \ref{fig:zz-lattice}).  As no spin-orbit terms are included, it suffices to consider a single spin species, and the spin index is therefore suppressed. The lattice constant has been set to unity.

	The model contains three types of hopping parameters, $t^a, t^b$, and $t^c$, associated with the three bond orientations of the honeycomb lattice (inset of Figure \ref{fig:zz-lattice}).  With the choice of unit cell shown, $t^a$ parameterizes hopping within the unit cell along the horizontal direction, $t^b$ describes hopping to the next unit cell along the horizontal direction, and $t^c$ represents hopping along the vertical direction within the unit cell. 
In general we will not require any lattice symmetries, such that all three types of hopping parameters depend explicitly on the indices $i$ and $j$.

\begin{figure}
		\centering
		\includegraphics[width=\linewidth]{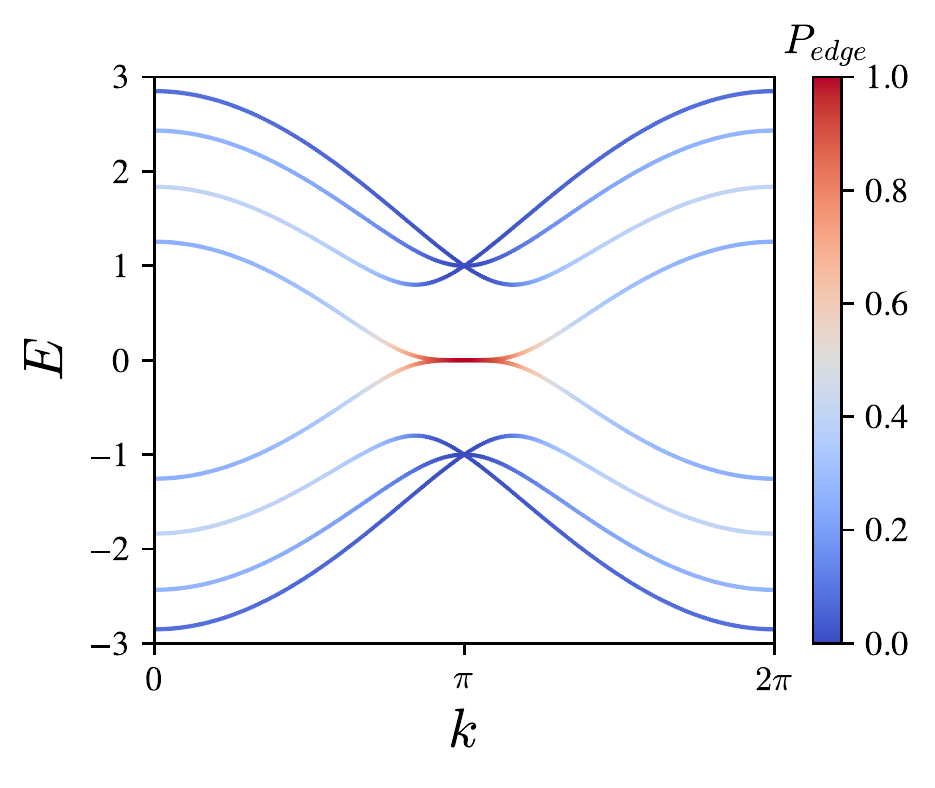}
		\caption{Band structure of the zigzag ribbon of width $W=4$ is shown. The bands are colored by the probability of the corresponding wavefunctions to be localized on the $A (B)$ sub-lattice on the bottom (top) of the zigzag ribbon, denoted by $P_{edge}$.}
		\label{fig:zz-dispersion}
	\end{figure}

	The clean limit of the zigzag ribbon is obtained by setting $t^{\alpha}_{i, j} \equiv t$.  The resulting band structure for  $W=4$ and $t=1$ is shown in figure \ref{fig:zz-dispersion}.   
	For a general width $W$, the clean system has $2(W-1)$ fully gapped bands, and one pair of gapless bands, which cross at $k=\pi$. Near this degeneracy point (approximately in the region $2 \pi/3 < k < 4 \pi/3$ spanning the projection of the bulk Dirac cones), the dispersion is very flat, and  the wave-function is localized near the ribbon's zigzag edges, as indicated by the coloring of the bands in figure \ref{fig:zz-dispersion}. In contrast, the wavefunctions of the fully gapped bands, as well as of the gapless band far from the degeneracy point, are delocalized throughout the bulk.  The Dirac points of 2D graphene (located at $k = 2 \pi/3$ and $4 \pi/3$ in the limit $W \rightarrow \infty$)  are separated from zero energy by a finite-size gap that scales as $1/W$. Hereafter we always focus on energies smaller than this $1/W$ gap, where only the two flat sub-bands are present.

\begin{figure}
		\centering
		\includegraphics[width=\linewidth]{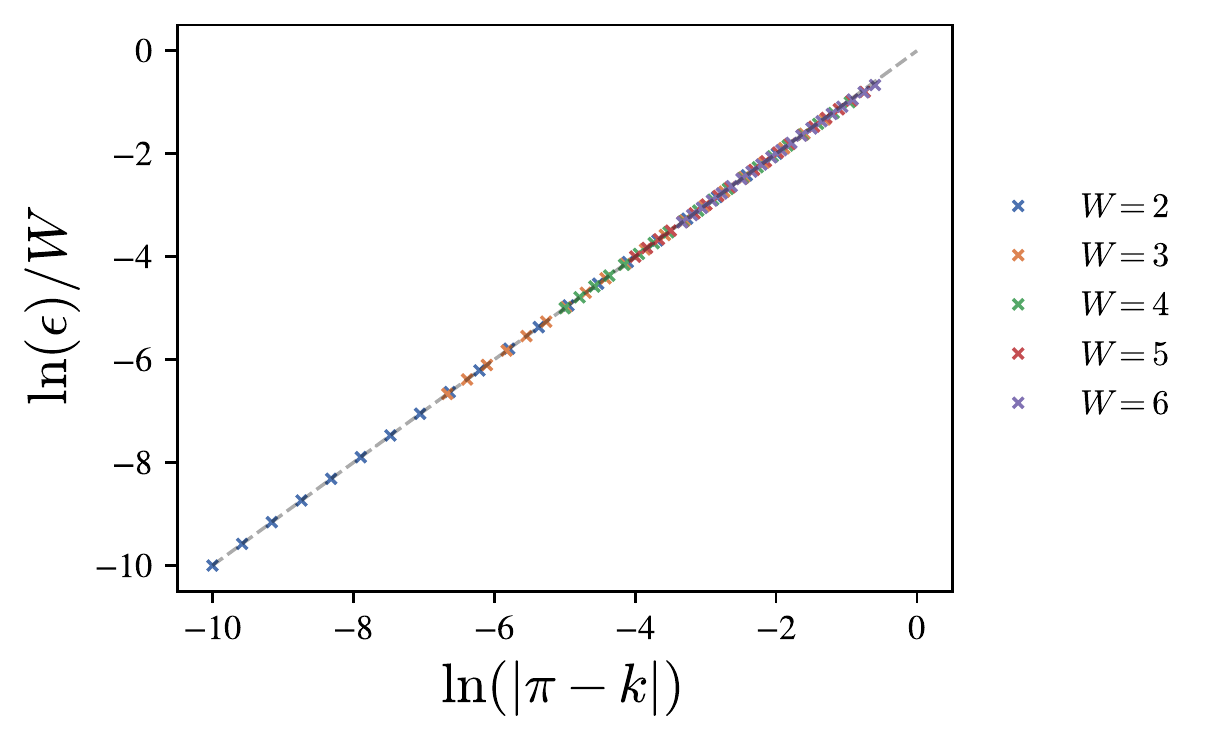}
		\caption{The low energy band as a function of momentum close to the degeneracy point. The axes are rescaled so that all the curves fall on the same dashed gray line, in agreement with Eq.~\ref{eqn:zz-low-energy-dispersion}.}
		\label{fig:zz-dispersion-low-energy}
	\end{figure}

	The analytical solutions for the energies and wave functions of clean zigzag ribbons can be found in Appendix B of Ref. \cite{WakabayashiGrapheneRibbons}.  A notable feature of the wave functions of the edge states that make up the lowest energy band  is that they are sub-lattice polarized, with support only on sub-lattice $A$ ($B$) of the honeycomb lattice at the lower (upper) edge. Upon expanding the dispersion of this band near the band crossing at $k=\pi$, one finds that:
	\begin{equation}\label{eqn:zz-low-energy-dispersion}
		\epsilon(k) \approx \pm \abs{\pi-k}^{W}.
	\end{equation}
	At energies below the finite-size gap, this is in a good agreement with the dispersion obtained via numerical diagonalization, as shown in figure \ref{fig:zz-dispersion-low-energy}.

	\subsection{Symmetry and topology}\label{subsec:topology}

	\subsubsection{Generic symmetries of zigzag graphene}

	Before discussing disordered ribbons, it is crucial to understand the symmetries of the Hamiltonian (\ref{eqn:ham-zz-real-space}), as well as the band structure topology of the clean zigzag ribbons discussed above. 
	
	Systems of non-interacting fermions can be classified into 10 distinct symmetry classes based on the existence of three generic symmetries \cite{AltlandZirnbauer,RyuTenFoldWay}: time-reversal ($T$), particle-hole ($C$), and chiral or sub-lattice ($S$). These are best understood by examining the so-called first quantized Hamiltonian $\mathcal{H}$, where $H = \sum_{\alpha,\beta} c^\dagger_\alpha c_\beta \mathcal{H}_{\alpha, \beta}$, and $\alpha,\beta$ label individual sites in the system. Time-reversal symmetry requires that:
	\begin{equation}
		U_T^\dagger \mathcal{H}^* U_T = \mathcal{H},
	\end{equation}
	where $U_T$ is a unitary operator. In the case at hand, the fermions are effectively spinless, and $U_T$ is the identity matrix, as it maps a site to itself.    It follows that the Hamiltonian (\ref{eqn:ham-zz-real-space}) has time-reversal symmetry, since all the hoppings are restricted to real values, and that $T^2 = 1$.
	
	Chiral symmetry is present if there exists a unitary matrix $U_S$  such that 
	\begin{equation}
		U_S^\dagger \mathcal{H} U_S = -\mathcal{H} .
	\end{equation}
	When $\mu=0$, any free fermion Hamiltonian on a bipartite lattice for which hopping terms connect only the $A$ and $B$ sublattices has chiral symmetry.  Explicitly, 
	\begin{equation}
		(U_S)_{\alpha,\beta} = 
		\begin{cases}
			0, & \alpha \neq \beta \\
			1, & \alpha=\beta \text{ belongs to sub-lattice A} \\
			-1, & \alpha=\beta \text{ belongs to sub-lattice B}
		\end{cases}.
	\end{equation}
	Thus, because we take all $t^\alpha_{ij}$ to be real, the Hamiltonian (\ref{eqn:ham-zz-real-space}) has both time reversal and chiral symmetries.  
	Since $S = T C$, particle-hole symmetry is also present and squares to $+1$.  As a result, the zig-zag nanoribbons studied here, for which $\mu = 0$ and there is no chemical potential disorder, belong to symmetry class BDI.

	\subsubsection{Zigzag graphene as a multi-critical point}

	In one dimension, symmetry class BDI admits topologically non-trivial  gapped phases, characterized by an integer-valued topological invariant, the winding number \cite{RyuTenFoldWay,KitaevClassification,LudwigClassification}. For the hopping models studied here, such gapped phases can be obtained by introducing staggered hopping parameters. For a single chain $(W=1)$, taking $t^a_{ij} \equiv  t^a, t^b_{ij} \equiv t^b$ in Eq. (\ref{eqn:ham-zz-real-space}) gives the Su-Schrieffer-Heeger (SSH) model \cite{SSH, ZakPhase}, which describes a 1D topological insulator in symmetry class BDI with winding $1$ ($t^b > t^a$) or $0$ ($t^b < t^a$).  The limit $t^a=t^b$ corresponds to a critical point separating these two phases.

\begin{figure}
		\centering
		\includegraphics[width=\linewidth]{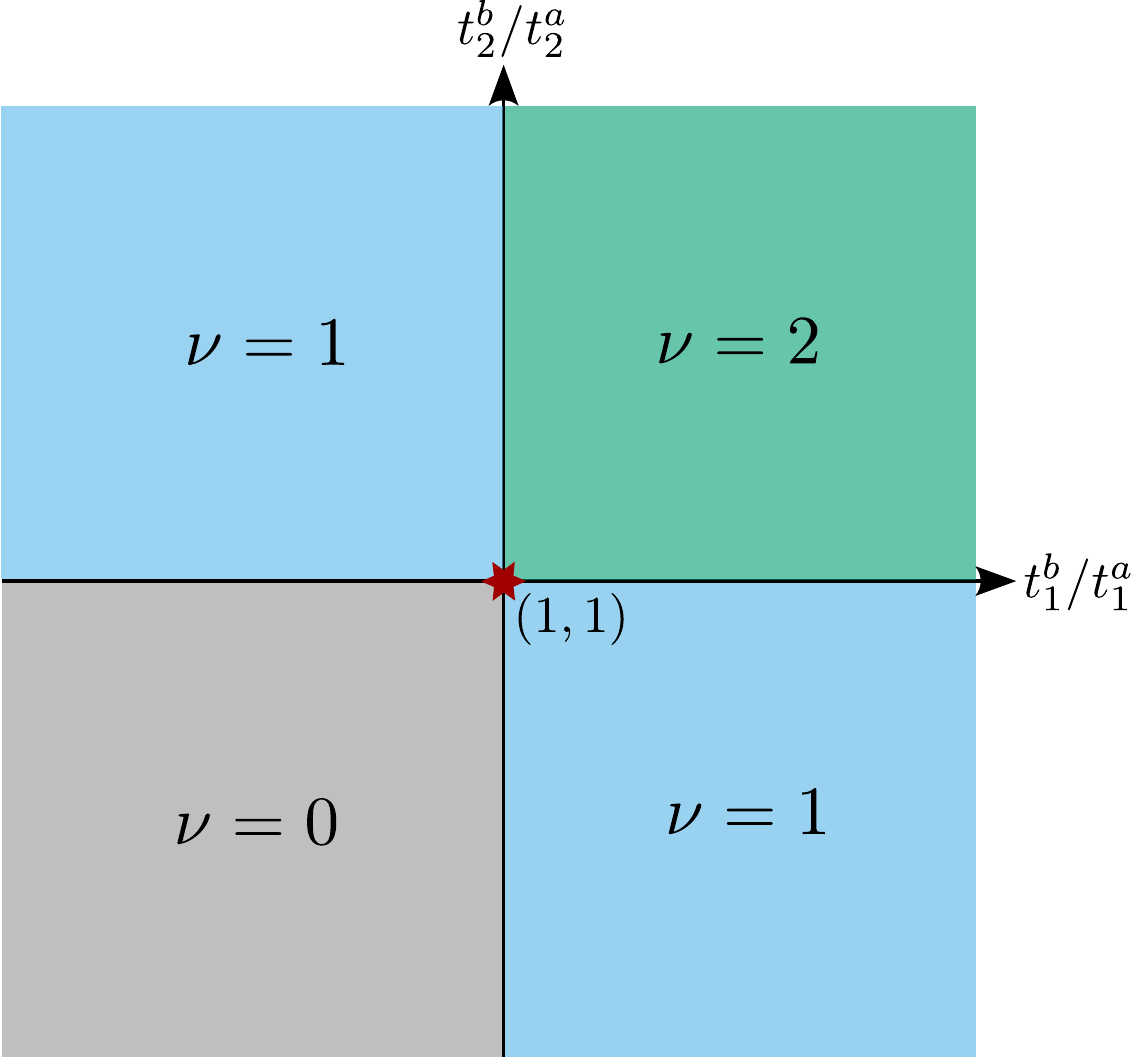}
		\caption{Phase diagram for the $W=2$ zigzag ribbon as a function of the hopping parameters, when there is no disorder. $\nu$ is the winding number at half-filling. The vertical hopping parameter $t_1^{c}$ is not relevant to the topology, in accordance with Eq. \ref{eqn:winding-number-width-2}.}
		\label{fig:phase-diagram}
	\end{figure}

	For $W=2$,  with translation invariance there are 5 distinct hopping parameters $t_{j}^{a}, t_j^b$ ($j=1,2$), and $t_1^c$. (Here the index $i$ labeling the horizontal unit-cell is omitted due to translation invariance). The Hamiltonian can be written in momentum-space as:
	\begin{equation}
		H = \sum_{k} \Psi^\dagger_k \mathcal{H}(k) \Psi_k,
	\end{equation}
	where $\Psi^\dagger_k = \begin{pmatrix} c^\dagger_{k, 1, A}	& c^\dagger_{k, 2, A} & c^\dagger_{k, 1, B} & c^\dagger_{k, 2, B}\end{pmatrix}$. This choice of basis is most convenient to find the winding number, since the Hamiltonian in momentum space is block off-diagonal \cite{RyuTenFoldWay,MatsuuraGaplessTopology}, i.e.:
	\begin{equation}
		\mathcal{H}(k) = \begin{pmatrix}
			0 & D(k) \\
			D^\dagger(k) & 0
		\end{pmatrix},
	\end{equation}
	where:
	\begin{equation}
		D(k) = \begin{pmatrix}
			t_1^a + t_1^b e^{-i k} & 0 \\
			t_1^c & t_2^a + t_2^b e^{-i k}
		\end{pmatrix}.
	\end{equation}
	The winding number is given by \cite{RyuTenFoldWay,MatsuuraGaplessTopology}:
	\begin{equation}
		\nu = \frac{i}{2\pi} \int_{0}^{2\pi} d k \Tr \left[ D_k^{-1} \partial_k D_k \right].
	\end{equation}
	One then finds:
	\begin{equation}
		\nu = \theta\left(\abs{t_1^b/t_1^a} - 1\right) + \theta\left(\abs{t_2^b/t_2^a} - 1\right),
		\label{eqn:winding-number-width-2}
	\end{equation}
	where $\theta$ is the step function. This is simply the sum of the winding numbers of the individual horizontal 1D chains. Notably, the vertical hopping between the chains does not play a role in the topology of the $W=2$ zigzag ribbon.  Taking $t_1^\alpha = t_2^\alpha = t^\alpha$ yields a winding number of $2$ ($t^b > t^a$) or $0$ ($t^b < t^a)$.  As before, the gapless model corresponding to uniform hopping parameters $t^a = t^b = t^c$ describes the critical point separating the two phases with winding numbers $0$ and $2$. The other possibilities for the topology of the $W=2$ ribbon are summarized in figure \ref{fig:phase-diagram}.

	This finding can be generalized for larger widths. In general, a translationally invariant ribbon of width $W$ can be viewed as an array of $W$ chains coupled via the vertical hoppings $t^c_j$, $j=1 ... W-1$, each with two intra-chain hopping parameters $t^a_j, t^b_j$. When $t_j^a = t_j^b$ for all $j$, the system is gapless, with a dispersion near the band-crossing that goes as $\abs{\pi-k}^W$. If all the chains have $t_i^a \neq t_i^b$, the spectrum is fully gapped with a winding number ranging anywhere from $0$ to $W$, given by
	\begin{equation} 
		\nu = \sum_{j=1}^W \theta(m_j), 
		\label{eqn:winding-W}
	\end{equation} 
	where $m_j=|t^b_j|-|t^a_j|$ is the SSH staggering along the $j$-th chain. Thus the clean limit discussed above corresponds to a multicritical point separating gapped topological phases of windings $0$ and $W$.

	A continuum low energy theory of the edge mode can be represented by an effective 1D Hamiltonian of the form:
	\begin{equation}
		H= \left(\begin{array}{cc}
		\hskip -.7cm 0&  \hskip -1.4cm (\partial_x + m_1)\ldots (\partial_x + m_W)\\
		(-\partial_x + m_W)\ldots (-\partial_x + m_1) & 0
		\end{array} \right)
		\label{eqn:continious-ham}
	\end{equation}	
where the $2\times 2$ structure is associated with $A/B$ sub-lattices. 	 
For $x$-independent staggerings $m_j$, the corresponding spectrum is given by:
\begin{equation}
	\epsilon(p) =\pm\prod_{j=1}^W \sqrt{p^2+m_j^2},
	\label{eqn:continuum-dispersion}
\end{equation}
where $p=\pi-k$. This leads to the spectrum (\ref{eqn:zz-low-energy-dispersion}) in the uniform hopping limit, $m_j=0$. Moreover, it is clear from here that to open a gap in the spectrum, {\em all} chains must be staggered. If there are $w<W$ non-staggered chains, the low energy spectrum is gapless and scales as $\epsilon(p) \propto |p|^w$. It is also clear from Eq.~\ref{eqn:continious-ham} that  the corresponding winding number is given by Eq.~\ref{eqn:winding-W}. 

If one of the staggerings changes sign over the system's length, e.g. $m_i(x)=|m_i|\mathrm{sign}(x)$, while all others remain finite everywhere, there is a zero energy in-gap state localized to the B sub-lattice and given by the solution of:
\begin{equation}
	(\partial_x + m_i(x))(\partial_x + m_{i+1})\ldots (\partial_x + m_W)\psi_B=0.
\end{equation} 
This leads to an homogeneous equation for the unnormalized wavefunction $(\partial_x + m_{i+1})\ldots (\partial_x + m_W)\psi_B=e^{-|m_i| |x|}$, which for constant $m_{i+1},\ldots m_W$ may be easily solved by e.g. the Fourier transform.   On the other hand, if  $m_i(x)=-|m_i|\mathrm{sign}(x)$ the zero energy state, 
found from $(-\partial_x + m_{i-1})\ldots (-\partial_x + m_1)\psi_A=e^{-|m_i| |x|}$,
is localized to the A sub-lattice.

	\subsubsection{Zigzag graphene as a topological semi-metal}

	In addition to this, two-dimensional graphene can be thought of as a gapless topological material \cite{MatsuuraGaplessTopology}. The bulk band structure contains a pair of Dirac nodes located at the two distinct corners of the Brillouin zone. Through a bulk-boundary correspondence, the topological nature of the Dirac nodes arising from their non-trivial Berry phase of $\pm \pi$ gives rise to low-energy states localized on the boundaries. This is a 2-dimensional analog of the Fermi arcs that exist at the surface of a 3D Weyl semi-metal \cite{WeylFermiArc,WeylExperiment1,WeylExperiment2}.

	To see how the bulk-boundary correspondence plays out in the quasi-1D system, one can consider the momentum-dependent Hamiltonian $\mathcal{H}(k)$ of a ribbon, where $k$ is momentum along the translationally invariant direction. In the case of a zigzag ribbon, this is given by the $W\times W$ matrix:
	\begin{equation}
		\begin{pmatrix}
			0 & 1+e^{-i k} \\
			1+e^{i k} & 0 & 1 \\
			 & 1 & 0 & 1+e^{-i k} \\
			 & & 1+e^{i k} & 0 & 1 \\
			 & & & & \ddots \\
			 & & & & 1+e^{i k} & 0
		\end{pmatrix}.
	\end{equation}
	At fixed $k$, one can view this as a Hamiltonian of a $W$-site 1D system.  In the case at hand, $\mathcal{H}(k)$ is identical to the Hamiltonian of an SSH chain without time-reversal symmetry (class AIII), with $t^a = 1 + e^{-i k}$, and $t^b = 1$. This chain is topological when:
	\begin{equation}
		\abs{1+e^{-i k}} < 1,
	\end{equation}
	i.e when $2\pi/3 < k < 4\pi/3$. The endpoints of the topological region are exactly where the Dirac points occur in the limit $W \rightarrow \infty$. For the values of momenta where the chain is topological, there are states close to zero energy that are localized on the boundaries of this auxiliary1D system at fixed $k$ (i.e. at the zigzag edges). These are the edge states seen in the dispersion of figure \ref{fig:zz-dispersion}. The existence and dispersion of the edge states depends on the boundary condition of the ribbons \cite{WakabayashiGrapheneRibbons,GrapheneQuantumDotDisorder}. For instance, in the arm-chair ribbons, the Dirac points with opposite Berry phase are crossed simultaneously and thus no edge states occur.

	Because the topology in this case is explicitly related to the existence of the translation-invariant Hamiltonian $\mathcal{H}(k)$, it is not  clear if the edge states  persists at a finite disorder strength. (This is in contrast to gapped topological phases, where boundary states are known to be robust provided that all relevant symmetries are preserved). It is thus interesting study the stability of edge states in the presence of disorder that preserves the chiral symmetry (which is required for the existence of the edge states in the SSH chain).

	\section{Off-diagonal disorder in zigzag graphene} \label{sec:disorder}

	\subsection{Disordered critical point in the 1D chain: A Review}

	To provide context for understanding transport and density of states in disordered zigzag graphene ribbons, it is helpful to review the properties of the disordered 1D random hopping chain at the critical point.	A body of work \cite{MotrunichDamleHuse,Deloc1DChains,GruzbergSuperuniversality} has established that the critical point separating phases of different winding persists to finite disorder strength.  If the distribution of hopping parameters is the same on each bond, adding disorder in the un-staggered chain leads to an unconventional type of Anderson localization, in which the average conductance at zero energy decays as a power of the system size, rather than exponentially. The corresponding transport behavior is described by the Fokker-Planck equation in symmetry class BDI \cite{Deloc1DChains,BrouwerDirtySC,RyuCrossover}, which for a single propagating channel leads to the following transport distribution :
	\begin{equation}
		P(x; s) = \sqrt{\frac{2}{\pi s}} e^{-\frac{x^2}{2s}}; \quad x \geq 0
		\label{eqn:bdi_prob}
	\end{equation}
	Here $s = L/l$, where $L$ is the system size, $l$ is the mean free path, and in the case of interest here:
	\begin{equation}
	g = \frac{1}{\cosh[2](x)} \ .
	\end{equation}
	To see that this distribution describes unconventional localization, one can define the typical and average localization lengths: 
	\begin{equation}
		\xi_{typ} \equiv -\frac{2 L } {\expval{\ln\left(g\right)}}  , \ \ \  \xi_{avg} \equiv - \frac{2 L}{ \quad \ln\left(\expval{g}\right) }.
		\label{eqn:xi}
	\end{equation}
	For the distribution (\ref{eqn:bdi_prob}), one finds $\xi_{typ} \sim \sqrt{Ll}$ and $ \xi_{avg} \sim L/ \ln s$, indicating that both typical and average localization lengths diverge with the system size.  In contrast, for conventional Anderson insulators, say in class AI, one finds that $\xi_{typ} = 2 l$ and $\xi_{avg} = 8 l$.

	Since the clean zigzag ribbon with homogeneous hoppings is at the multicritical point, one expects that this criticality will persist in the presence of hopping disorder.  This suggests that adding disorder will lead to unconventional localization, much like in the disordered hopping chain. However, the precise nature of this unconventional localization could be very different.

	For ribbons of width $W>1$, an additional subtlety arises in trying to determine the nature of the underlying critical point.  For $W=1$ (or in the case of armchair  nanoribbons \cite{TwoParamScaling}), one can deduce the presence of the critical point by considering the conductance of a disordered segment connecting two disorder-free leads, when the incident particle has energy $\epsilon = 0$.  For $W \geq 2$, however, the transport at zero energy cannot be accessed directly, because the velocity of the band goes to zero at $\epsilon = 0$. Therefore the injected particle does not propagate through the disordered region. Instead, the nature of the critical point at zero energy must be inferred by studying transport at small but finite energies.

	This difference is significant because the distribution (\ref{eqn:bdi_prob}) characteristic of critical transport in symmetry class BDI is valid only when the energy is exactly zero, where chiral symmetry imposes additional constraints on the allowed scattering processes.  From the transport point of view, moving away from zero energy is therefore analogous to tuning away from the critical point.  For a fixed energy and disorder strength, this leads to a crossover between transport reminiscent of the chiral distribution (\ref{eqn:bdi_prob}) at shorter length-scales, to more conventional localization and exponentially suppressed conductance at long distances. For $W \geq 2$, the nature of the underlying critical point must be inferred by studying transport in this crossover regime.

	The study of this crossover dates back to work by Dyson\cite{DysonDOS}.  A key observation is that when the clean system is at a critical point between phases of different winding, disorder creates domains of each of the corresponding phases.  The resulting domains walls harbor bound states, whose energy is exponentially small in the domain size.  These low-energy states are thus resonant and can hybridize, leading to delocalization  \cite{MotrunichDamleHuse}. When the disorder distribution is uniform (i.e. no net staggering), such domains form on all length scales, and the eigenstates of the disordered system are not exponentially localized even at the longest length scales, as indicated by the divergence of both typical and average localization lengths.

	This intuitive picture explains several of the unusual features seen in disordered 1D systems  proximate to a topological phase transition.  First, as pointed out by Dyson \cite{DysonDOS,TitovDOS,BrouwerFPReview},
	the density of states diverges at low energies according to: 
		\begin{equation}\label{eqn:dyson-dos}
			\rho(\epsilon) \propto  \frac{1}{\abs{\epsilon (\ln \epsilon)^3}}.
		\end{equation}
	When the system is perturbed away from the critical point by, say, introducing staggering of the bonds on average, this divergence becomes a power-law $\epsilon^{-1+\delta}$ with a non-universal exponent $\delta$  \cite{MotrunichDamleHuse}.

	Second, as the energy decreases towards zero, the typical localization length defined in Eq. \ref{eqn:xi} has a characteristic logarithmic divergence: \cite{Theodorou1DChain, EggarterLocDivergence, ZimanLocDivergence, DFisherIsingChains, McKenzieQPT,Kamenev2014}	
	\begin{equation}\label{eqn:xi-divergence}
			\xi_{typ} = l \abs{\ln(r)}.
	\end{equation}
	Here $r = \epsilon \tau$ is a dimensionless measure of the energy, $\epsilon$ is the energy of the injected electron, and $\tau$ the scattering time, and the formula is valid for $r\ll1$.  Moreover, as zero energy is approached, transport becomes increasingly dominated by rare-region effects and one finds that the {\it average} localization length diverges as $\log^2(r)$ \cite{BalentsFisher}.  Evidently, for finite $L$, both divergences are cut off at sufficiently low energies by the $L$-dependent localization lengths of the zero-energy BDI case.

	Finally, the full transport statistics of the disordered $W=1$ chain near zero-energy were studied in Ref. \cite{TwoParamScaling}.   In that case, the distribution of conductance values, and therefore all transport properties, is determined by the two dimensionless parameters $s = L/l$, and $r=\epsilon \tau$.  When $s \gg 1$, or $r \ll 1$, the transport distribution was found to be of the form:
	\begin{equation}
		P(x) = \frac{2}{\gamma \, \Gamma(\delta/2)} \left( \frac{x}{\gamma}\right)^{\delta-1} \exp\left[- \left( \frac{x}{\gamma} \right)^2 \right], \ x\geq 0
		\label{eqn:gamma_dsbn}
	\end{equation}
	where $\delta$ and $\gamma$ are functions of $s, r$.  The parameter $\delta$ dictates the shape of the distribution: 
	While the chiral distribution \ref{eqn:bdi_prob} is obtained for $\delta = 1$, a Gaussian distribution corresponding to exponential localization is obtained as $\delta \rightarrow \infty$.  For $r \ll 1$, one finds
	\begin{equation}
	\delta \approx \frac{3}{2} \frac{s}{|\ln^2 r | } \ ,
	\end{equation}
	indicating that the overall shape of the distribution is controlled by the ratio of $\xi_{\text{typ}}^2/l \sim \xi_{\text{av}}$ to the system size $L$.

	\subsection{Disordered zig-zag graphene nanoribbons: theoretical expectations} \label{PertSec}

	In the remainder of this section, numerical evidence is presented to indicate that for a range of widths $W$, disordered zigzag ribbons exhibit the same 2-parameter scaling with $s$ and $r$ as the 1D chain, with an underlying distribution of the same form. Because the dispersion of the disorder-free critical point depends strongly on $W$, this nevertheless describes a family of disordered models with quantitatively very different transport properties.

	To set the stage for the numerical results that follow, it is instructive to consider how the relaxation time $\tau$ and the mean-free-path $l$ might depend on the width $W$ and energy $\epsilon$. This can be done analytically when the disorder is sufficiently weak, such that  the scattering rate can be calculated perturbatively using Fermi's golden rule, which gives:
	\begin{equation} \label{eqn:fermi-golden-rule}
		\tau^{-1}(\epsilon) = 2 \pi \rho_0(\epsilon) \abs{\bra{\psi_L} H_{imp} \ket{\psi_R}}^2.
	\end{equation}
	Here $\rho_0$ is the unperturbed density-of-states, $\ket{\psi_L} (\ket{\psi_R})$ is the wavefunction for the left (right) moving mode, and $H_{imp}$ is the impurity Hamiltonian.

	The right-hand side of Eq. (\ref{eqn:fermi-golden-rule}) contains two terms that depend on the width of the ribbon. First, the dispersion of the clean system leads to the density of states 
	\begin{equation}
		\rho_0(\epsilon) = \frac{1}{\pi W} \epsilon^{-1+1/W},
		\label{eqn:DOS-clean}
	\end{equation} 
	which for $W >1$ has a power-law divergence as $\epsilon \rightarrow 0$.  
	Second, the matrix elements $\bra{\psi_L} H_{imp} \ket{\psi_R}$ also depend 
	on $W$.  To see why, consider a single impurity located on unit-cell $0$ and chain $m$ with a strength of $V$, i.e.:
		\begin{equation}
			H_{imp} = V c^\dagger_{0, m, A} c_{0, m, B} + \text{h.c.}
		\end{equation}
	Using the analytical form of the wavefunctions of the edge states in Ref. \cite{WakabayashiGrapheneRibbons}, and expanding about $\epsilon = 0$, gives a scattering amplitude $\bra{\psi_L}H_{imp}\ket{\psi_R}\sim V \epsilon^{1-1/W}$ which vanishes as $\epsilon \rightarrow 0$.  This strong suppression of back-scattering results from the fact that the edge states are well localized on the $A$ sub-lattice near the bottom edge and the $B$ sub-lattice near the top edge (see section \ref{subsec:model-and-spectrum}), such that backscattering between the left and right moving states is highly suppressed in the presence of off-diagonal disorder.

\begin{figure}
		\centering
		\includegraphics[width=\linewidth]{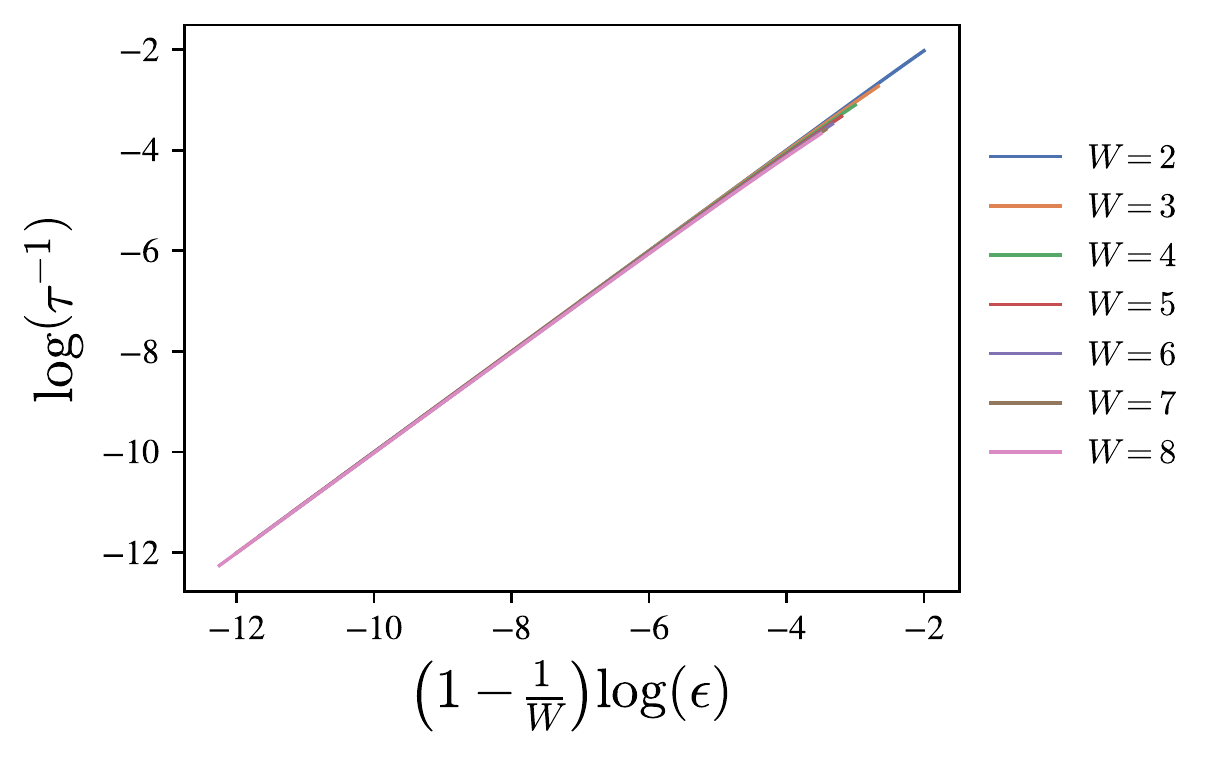}
		\caption{The relaxation rate calculated numerically using the Fermi's golden rule is plotted as a function of the energy for several widths. The axes are chosen in such a way that all the curves collapse onto a single line according to Eq.~\ref{eqn:relaxation-rate-scaling}}
		\label{fig:relaxation-rate-scaling}
	\end{figure}

	The exact matrix elements can be easily computed using the right and left-moving wavefunctions obtained through numerical diagonalization. After averaging over all possible values of $m$, the relaxation rate scales as:
	\begin{equation}\label{eqn:relaxation-rate-scaling}
		\tau^{-1} \propto V^2 \epsilon^{1-1/W},
	\end{equation}
	in accordance with the analytical estimate described above. This is confirmed by the scaling plot in figure \ref{fig:relaxation-rate-scaling}.  The mean-free-path is given by $l = \tau v(\epsilon)$ where $v(\epsilon) \sim W \epsilon^{1-1/W}$ is the velocity.  Because of the non-linear dispersion, the Fermi velocity rapidly approaches zero as $\epsilon$ vanishes, giving:
		\begin{equation}\label{eqn:l-scaling}
			l \propto \frac{W}{V^2}.
		\end{equation}
	Thus the  energy-dependence of the velocity exactly cancels that of the relaxation rate, and the mean-free-path $l$ is roughly independent of energy, and increases linearly with the width $W$.

	\subsection{Numerical Results for transport in zigzag graphene}

	In order to study transport in zigzag ribbons, a finite size tight binding system of length $L$ and width $W$ is created with the Hamiltonian given by Eq.~\ref{eqn:ham-zz-real-space}. In the absence of disorder, all the hoppings are set to $1$. To add disorder, a random term from the uniform distribution spanning $[-V, V]$ is independently added to every hopping. The quantity $V$ is a measure of the disorder strength. Semi-infinite, disorder-free leads of the same width are attached on either side of the disordered strip. One can then send in electrons at a certain energy and compute the S-matrix, which relates the amplitudes of the incoming waves to the outgoing waves. From this the conductance and other transport properties can be obtained. All of this is done using methods available in the package KWANT for python \cite{KWANT}, which uses the MUMPS library to efficiently solve sparse linear equations \cite{MUMPS}.

	A range of widths $W$ from $2$ to $8$ are studied, with lengths $L$ ranging from 100 to 1000 unit cells. The disorder strength $V$ is taken to be $0.5$ for all the cases, unless stated otherwise. The energy of the incoming electron $\epsilon$ ranges from $10^{-6}$ to $10^{-2}$. Over $10^{4}$ disorder configurations are generated for each $W$ and $L$ and $\epsilon$ in order to obtain comprehensive statistics.

\begin{figure}
		\centering
		\includegraphics[width=\linewidth]{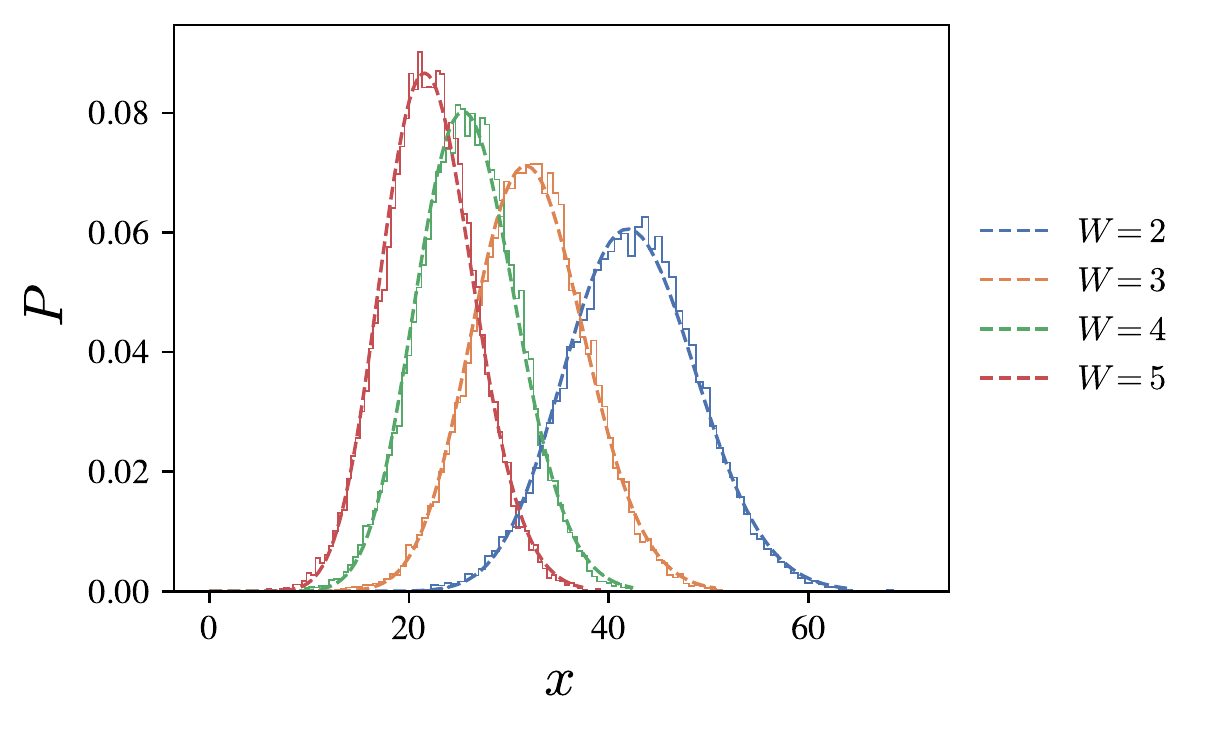}
		\caption{Probability distribution for $x=\mathrm{arccosh}(1/\sqrt{g})$ is shown for a range of widths for $L=1000$ and $\epsilon=10^{-2}$. The solid lines are the histograms obtained from numerics while the dashed lines are fits to Eq.~\ref{eqn:gamma_dsbn}.}
		\label{fig:prob_dsbn}
	\end{figure}

	As a first step, one can try to compare the numerically computed transport data of the zigzag chains to the transport statistics obtained for the 1D chain in Ref. \cite{TwoParamScaling}. In fig. \ref{fig:prob_dsbn}, it is shown that the probability distribution for the quantity $x = \mathrm{arccosh}(1/\sqrt{g})$ is well described by that of the 1D chain, Eq.~\ref{eqn:gamma_dsbn}. This suggests that the full transport statistics in a zigzag ribbon could be described by the same two-parameter scaling functions, with the parameters being $s=L/l$ and $r=\epsilon\tau$.  

\begin{figure}
		\centering
		\includegraphics[width=\linewidth]{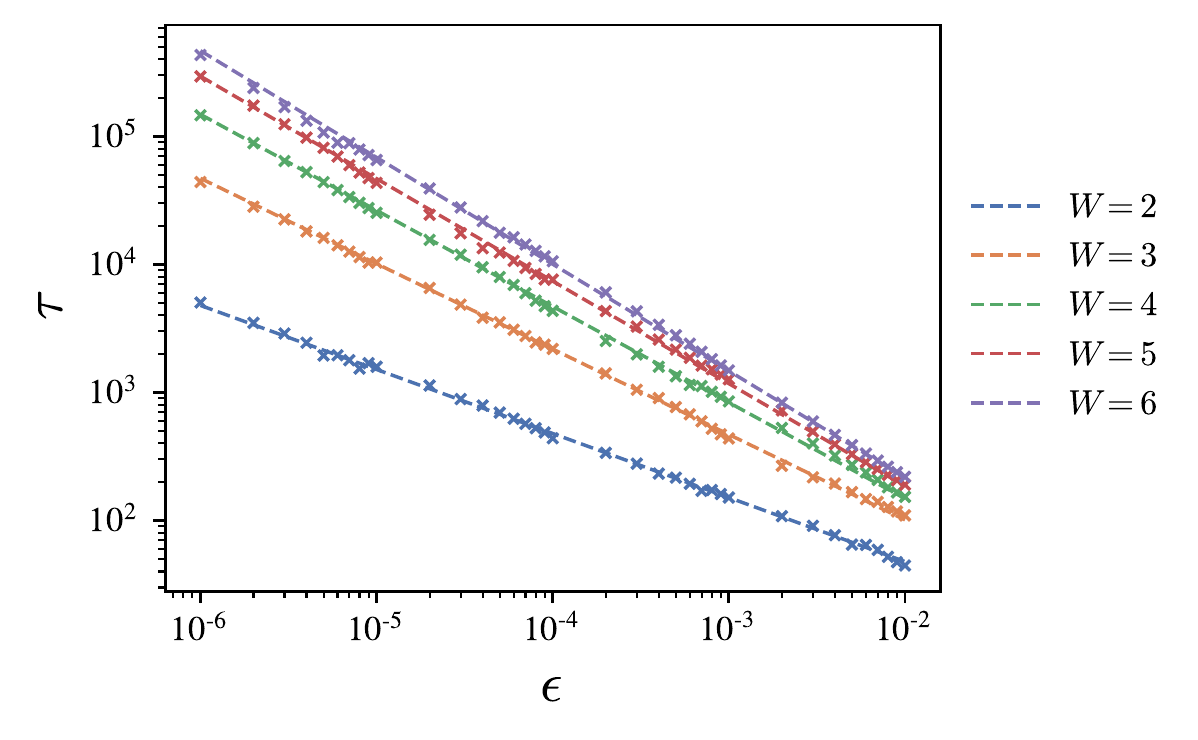}
		\caption{The scattering time $\tau$, obtained from fits of $\expval{\ln(g)}$ vs $L$, is shown as a function of energy for several widths. Based on Eq.~\ref{eqn:relaxation-rate-scaling}, the dashed lines show fits to $\tau = c \epsilon^{-1+1/W}$, where the only free parameter is $c$. In general, $c$ depends on the disorder strength. In this case it is around $1.5$ for all the lines. The fits show good agreement with the scaling relationship predicted by perturbation theory.}
		\label{fig:tau-vs-energy}
	\end{figure}

In order to study this scaling further, the values of $s$ and $r$ for a given zig-zag chain must be obtained.  In principle, this could be done by fitting to the distribution (\ref{eqn:gamma_dsbn}); in practice, however, it is more straightforward to fit the data of 
the zig-zag nanoribbons to that of 1d chain, for which these parameters can be extracted from the transport statistics at $\epsilon = 0$.  Details of this fitting procedure are given in appendix \ref{app:fitting}.  
The resulting energy and width dependence of the relaxation time $\tau$ is shown in Figure \ref{fig:tau-vs-energy}, and agrees strongly with both the energy and width dependence predicted by Eq.~\ref{eqn:relaxation-rate-scaling}. Similarly, the mean-free-path $l$ is also found to scale linearly with the width $W$ as predicted by Eq.~\ref{eqn:l-scaling} (this is not shown). To summarize, fits of the transport data of the zigzag ribbon to those of 1D chain, obtained by assuming the same two-parameter scaling of transport, show good agreement with perturbation theory.

\begin{figure}
		\centering
		\begin{subfigure}{\linewidth}
			\centering
			\includegraphics[width=\linewidth]{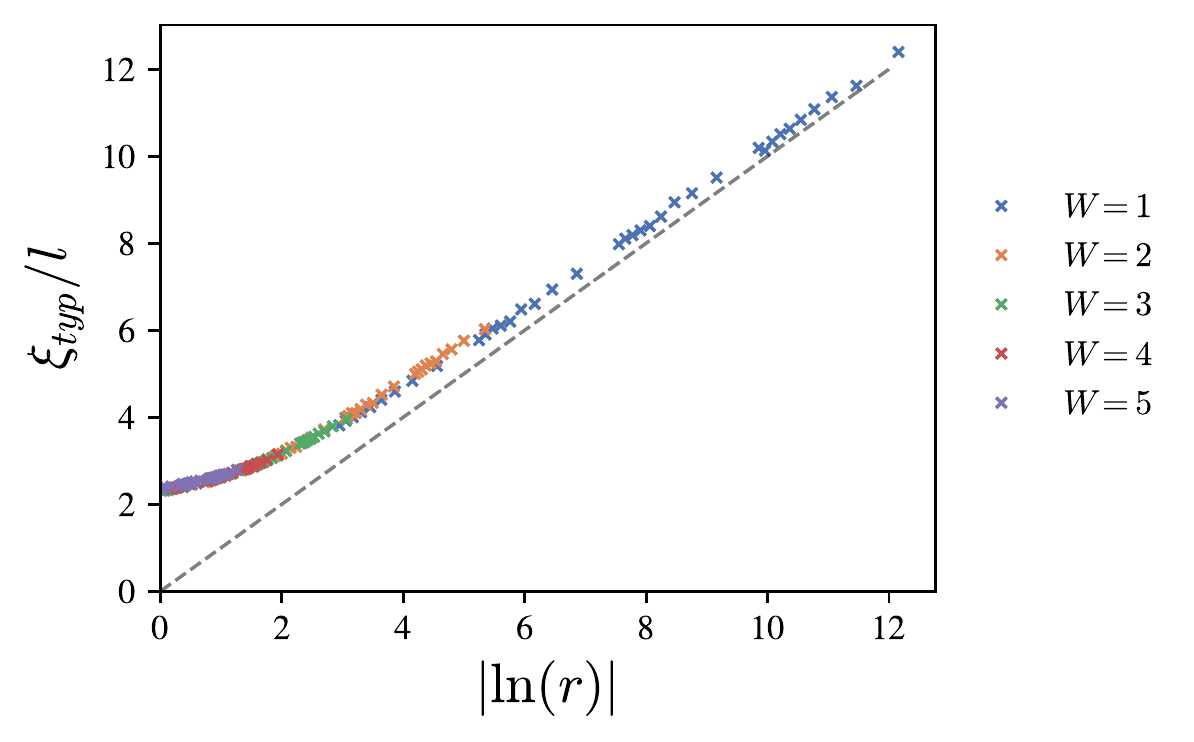}
			\caption{}
		\end{subfigure}
		\vfill
		\begin{subfigure}{\linewidth}
			\centering
			\includegraphics[width=\linewidth]{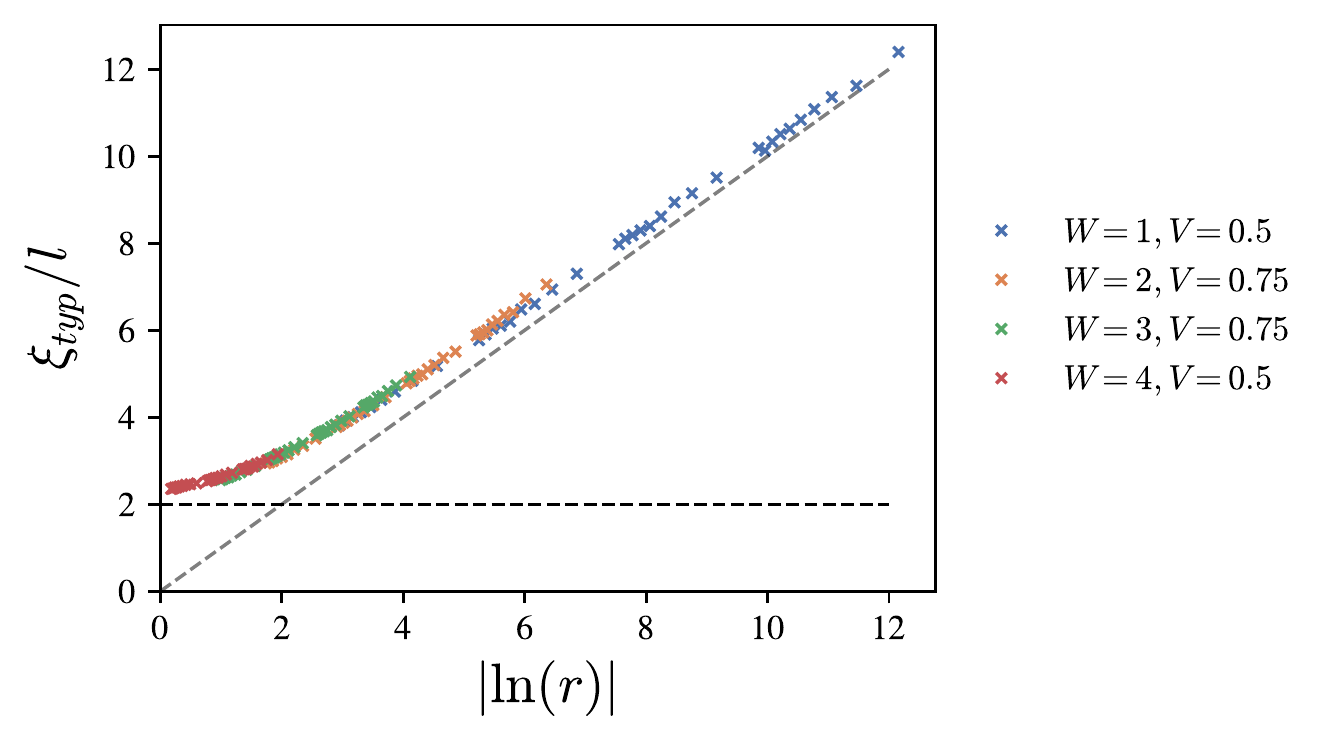}
			\caption{}
		\end{subfigure}
		\caption{The ratio of typical localization length to mean free path $\xi_{typ}/l$ is shown vs. $\abs{\ln(r)}$. Excellent data collapse is observed for several widths. Here $\xi_{typ}$ is obtained using Eq.~\ref{eqn:xi} and $r$ is obtained from fits to the data for the 1D chain (see appendix \ref{app:fitting}). The dashed line shows $\xi_{typ}/l = \abs{\ln(r)}$, which is the behavior expected for $r \ll 1$. Data for $W=1$ is included to show that logarithmic divergence is only attained when $\abs{\ln(r)} \gtrsim 5$. An additional disorder strength for a dew widths are included in (b) to show that the data collapse is not coincidental.}
		\label{fig:xi-vs-lnr}
	\end{figure}

	The agreement shown in Fig. \ref{fig:tau-vs-energy} suggests that the disordered  critical point in graphene nanoribbons exhibits the same scaling behavior as that of the random hopping chain.  A number of other measures also indicate the universality of this scaling. 
	Figure \ref{fig:xi-vs-lnr} shows that the typical localization length shows good data collapse for multiple widths when plotted as a function of $r$. A few features of this plot are worth emphasizing. For $\abs{\ln(r)} \approx 0$, the ratio $\xi_{typ}/l$ approaches $2$. This is the value predicted by the Fokker-Planck equation for symmetry class AI \cite{AbrikisovSolution,BeenakkerRMT}, where chiral symmetry is absent.  The logarithmic divergence of Eq.~\ref{eqn:xi-divergence} is only reached for $\abs{\ln(r)} \gtrsim 5$. This regime becomes increasingly inaccessible for the zigzag ribbons as the width increases, even though the smallest energy studied for all the systems is $10^{-6}$. This is because backscattering is suppressed as the width of the zigzag ribbon  increases  (see Eq.~\ref{eqn:relaxation-rate-scaling}), so that the regime $r \ll 1$ occurs at smaller energies for larger $W$.  In practice, even for $W=2$ much of this regime is  numerically inaccessible.

	\subsection{Density of states}

	To obtain a second probe of critical scaling, the density of states is computed using the recursive green's function technique \cite{MacKinnonRGF}. This is done completely independent of the transport calculations. The self-energy of the lead is added to the initial green's function to stabilize the results. Also, an imaginary part of $10^{-2} \epsilon$ is added to the energy. The maximum length studied is $L=10^4$, with energies ranging from $10^{-10}$ to $10^{-1}$, and $V=0.5$. Once again, around $10^4$ disorder configurations are computed.

	In terms of the scaling variables here, the density of states in the $W=1$ chain diverges at low energies according to:
	\begin{equation}\label{eqn:dyson-dos-2}
		\rho(\epsilon) = \frac{\pi \rho_0(\epsilon)}{\abs{r \ln^3(r)}},
	\end{equation}
	where $\rho_0$ is given by Eq. \ref{eqn:DOS-clean}. For the zigzag ribbons at low energies, using $\tau = c \epsilon^{-1+1/W}$ and the expression for $\rho_0$ \ref{eqn:DOS-clean}, one finds:
	\begin{equation}
		\rho(\epsilon) = \frac{W^2}{ \abs{2 c \epsilon \ln^3(c^W \epsilon)}}.
	\end{equation}
Thus, up-to an overall pre-factor, the density of states at any $W$ exhibits the same Dyson singularity as the disordered critical point separating phases whose topological winding number differs by one. 

In the absence of disorder, in contrast, the low-density of states of the zigzag ribbons depends strongly on their width, diverging as $\epsilon^{-1+1/W}$ at small energies (see Eq.~(\ref{eqn:DOS-clean})). Nevertheless the low energy density of states of disordered ribbons is completely universal and $W$-independent. This feature, in combination with the long relaxation time of the zigzag ribbons, means that the excess density of states due to disorder becomes apparent only at extremely small energies.   This is similar to what was observed with the localization length divergence. 

\begin{figure}
		\centering
		\begin{subfigure}{\linewidth}
			\centering
			\includegraphics[width=\linewidth]{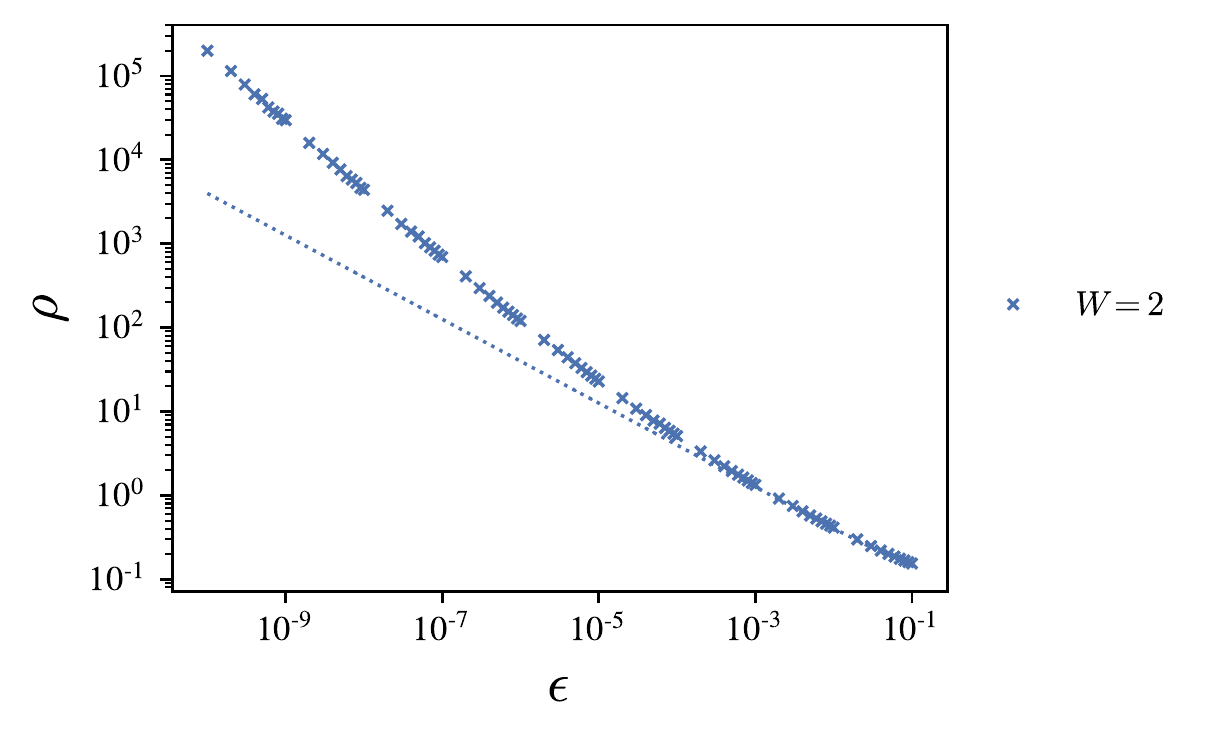}
			\caption{}
			\label{fig:dos-vs-energy}
		\end{subfigure}
		\vfill
		\begin{subfigure}{\linewidth}
			\centering
			\includegraphics[width=\linewidth]{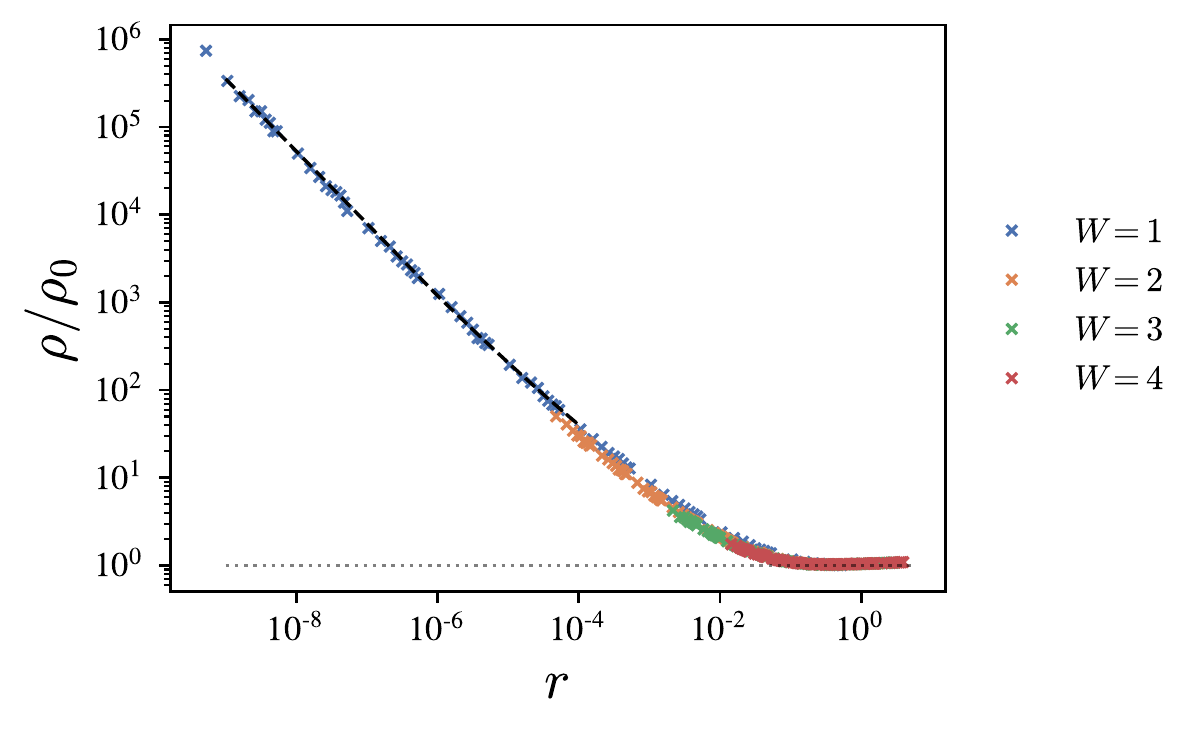}
			\caption{}
			\label{fig:dos-scaled-vs-r}
		\end{subfigure}\
		\caption{ The density of the states for disordered zigzag ribbons obtained using the recursive Green's function. Figure \ref{fig:dos-vs-energy} shows the density of states as a function of energy for a chain of width $2$, with the dotted line showing the density of states in the absence of disorder. The deviation from the dotted line becomes more pronounced at lower energies. Figure \ref{fig:dos-scaled-vs-r} shows the re-scaled density of states plotted as a function of $r = \epsilon\tau$. The gray dotted line shows the value in the absence of disorder, while the black dashed line corresponds to Eq.~\ref{eqn:dyson-dos}, which is valid for $r \ll 1$.}
		\label{fig:dos}
	\end{figure}

Figure  \ref{fig:dos} shows the low-energy divergence of the density of states obtained from numerics.  
In figure \ref{fig:dos-vs-energy}, the energy-dependent disordered density of states for $W=2$ is compared that of the clean system, showing that the Dyson singularity becomes detectable only at energies less than approximately $10^{-5}$ in this system.  Figure \ref{fig:dos-scaled-vs-r} shows the the density of states re-scaled by $\rho_0$ and plotted as a function of $r$.   Excellent data collapse is found across a range of widths. The observed $r$ dependence interpolates between the Dyson form of the divergence given by Eq.~\ref{eqn:dyson-dos} for $r \ll 1$ and the value at the clean limit for $r \sim 1$. Notably, the scaling parameter $\tau$ is obtained from the transport data, completely independent of the density of states calculations.

	With this, there is a sufficient evidence to conclude that the zigzag graphene ribbons are at a critical point with a delocalized state at zero energy, upon adding off-diagonal disorder.

	\subsection{Edge State Stability}

	Lastly, the robustness of the edge states to hopping disorder is studied. As noted above, when chiral symmetry is preserved, the existence of a Dyson singularity ensures that the density of states of the disordered system diverges as $\epsilon \rightarrow 0$.  The difference between chains with $W=1$ and $W\geq 2$ is that in the latter case, the low-energy density of states is divergent even in the clean system; this divergent density of states comes from the flat-band boundary modes.  
Moreover, since the low-energy modes of the clean system are localized to the edge, the matrix elements of the disorder Hamiltonian with the edge states are highly suppressed due to the sub-lattice polarized nature of the wave functions.  One might anticipate that these two effects combine to render the edge states effectively robust up to some finite disorder strength.  

\begin{figure}
		\centering
		\includegraphics[width=\linewidth]{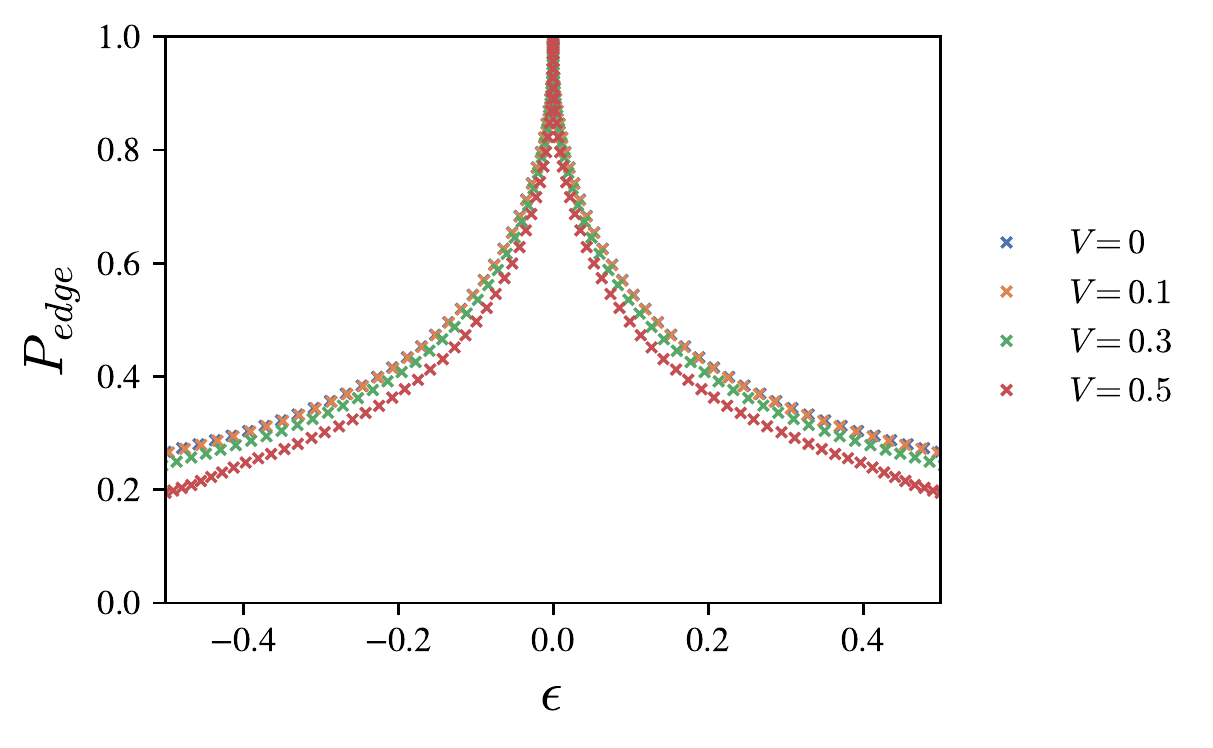}
		\caption{The probability of a state being localized on the zigzag edges $P_{edge}$ is shown as a function of the energy for a zigzag chain of width $5$ and a few disorder strengths. The persistence of the peak at zero energy indicates that the edge states are still present near zero energy.}
		\label{fig:edge-prob}
	\end{figure}

	To show that these expectations are indeed borne out, finite size zigzag ribbons of length $L=100$ and several widths are studied. The wave functions are calculated by exact diagonalization and $10^{3}$ disorder realizations are computed for each parameter value.	In figure \ref{fig:edge-prob}, the probability of a state being localized on either the A sub-lattice on the bottom of the chain, or the B sub-lattice on the top of the chain, denoted by $P_{edge}$, is shown for several disorder strengths and $W=5$. As seen in the Figure, disorder does not destroy the peak in the edge state probability at zero energy; rather it pushes the edge states closer to zero energy. This reflects the Dyson singularity. Moreover, disorder does not substantially alter the degree to which these low-energy states are localized to the system's boundaries, indicating that the edge states are stable with disorder.

\begin{figure}
		\centering
		\includegraphics[width=\linewidth]{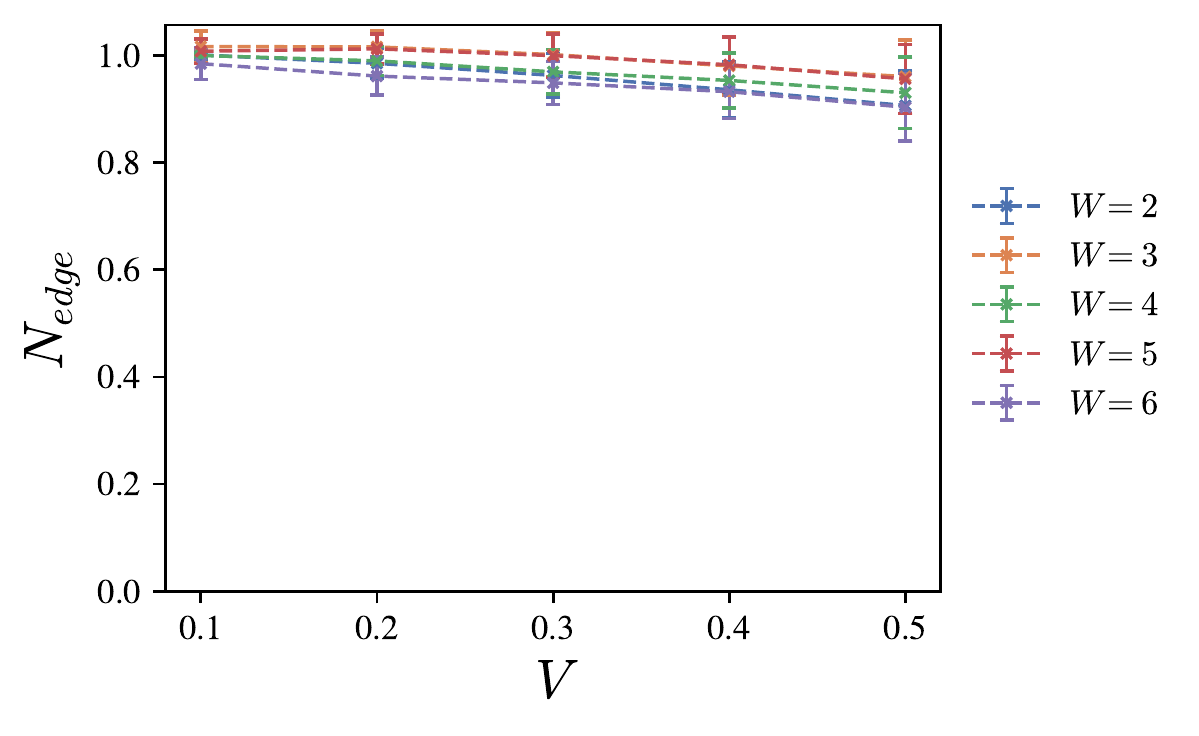}
		\caption{The ratio of number of edge states in a disordered system to those in the clean system $N_{edge}$, is shown as a function of disorder strength for a few widths. The number of edge states remains roughly the same upon increasing the disorder strength.}
		\label{fig:num-edge-vs-disorder}
	\end{figure}

	The total number of edge states can also be computed, by setting a threshold for $P_{edge}$ above which a state is considered an edge state.  Figure \ref{fig:num-edge-vs-disorder} plots the resulting  ratio of number of edge states in a disordered system to those in the clean system, with the threshold value for $P_{edge}$ taken to be $0.8$. The Figure shows that, for several different widths, this quantity varies very little as function of disorder strength, with the ratio of disordered to clean edge states very close to $1$ in all cases. This indicates that the edge states near zero energy persist upon increasing the disorder strength.  

	\section{Discussion}\label{sec:conclusions}

	Zigzag graphene with homogeneous nearest neighbor hopping is at a multicritical point between gapped topological phases identified by a topological winding number related to the width $W$ of the ribbon. As a result, it is particularly interesting to study these models with the addition of hopping disorder, which preserves the symmetry class BDI of the clean system. The resulting width $W$ zigzag ribbons can be viewed as an extension of the 1D SSH chain at criticality, where the disordered critical point is associated with the Dyson singularity in the density of states, and diverging localization length, at zero energy.

	Our work shows that the underlying crossover in transport statistics and the scaling of localization length and density of states are essentially the same as the critical 1D chain, after accounting for the difference in scattering time due to the slow velocities of the low-energy modes. 
The numerical evidence for this is substantial.  At non-zero energies near the critical point, the transport statistics of the 1D chain were previously found to obey a two-parameter scaling (Ref. \cite{TwoParamScaling}), where the two parameters $s = L/l$ and $r=\epsilon \tau$ completely determine any transport quantity. The transport data obtained for the zigzag ribbons of all widths studied here fits well to this two-parameter scaling data, with the relaxation times and mean-free-paths obtained from the fits in good agreement with the predictions from perturbative calculations.  Other quantities studied, such as the typical localization length, also show excellent data collapse across several widths, with evidence of the logarithmic divergence expected for the $W=1$ critical point seen for $r \ll 1$.  Moreover, the density of states is computed independently of the transport and also shows data collapse when plotted as a function of $r$.  All of this leads to the conclusion that there is indeed a disorder induced critical point at zero energy, obeying the same two-parameter scaling of transport at non-zero energy.

	Though this universal scaling collapse may seem to indicate that the multicritical point belongs to the same universality class as the ordinary disordered critical point in class BDI, this is actually not the case. The coincidence of the energy scaling of the localization length and the density of states is due to the fact that these scaling have a peculiar logarithmic character. This is {\em not} the case when considering other possible deviations from criticality.

	To be specific, let us consider a uniform staggering in all of the chains along the ribbon's direction. In  the Hamiltonian \ref{eqn:ham-zz-real-space}, this is achieved by choosing the hopping amplitudes such that $m_j=|t^b_j|-|t^a_j|$. Eq. \ref{eqn:continuum-dispersion} suggests that:
	\begin{equation}
		\xi_{typ} \propto \prod_{j=1}^{W} \frac{1}{\abs{m_j}^{\alpha}} .
		\label{eqn:xi-critical-exponent}
	\end{equation} 
A natural conjecture for  the exponent here is that it take on the value $\alpha=1$ \cite{Prodan2014,Kamenev2016} of the 1D chain.  To justify this, imagine taking $m_j \neq 0$ in all but one of the chains.  In this case there is a linear band crossing at zero energy (see Eq.~\ref{eqn:continuum-dispersion}), and the resulting critical point separates phases whose winding numbers differ by one, exactly as in the 1D SSH chain.  
It follows that for a uniform staggering, $m_j\equiv m$, the localization length scales as  $\xi_{typ} \propto m^{-W}$ and thus the corresponding 
	critical exponent is $W$ dependent.

	The unusual properties of the low energy band in zig-zag graphene ribbons follow from the fact that 2D graphene can be viewed as a topological semi-metal, which has edge states with zigzag boundary conditions.   These topological properties extend also to finite-width ribbons, and explain the nature of the low-energy edge states studied above. These edge states appear numerically to be stable in the presence of hopping disorder, remaining well-localized near the boundary, and very close to zero energy.  However, away from zero energy disorder does lead to localization in the transport of these wave-functions; thus at low energies the edge states remain present in the spectrum, but cease to be conducting.

%

	\section{Acknowledgements}

	This work was supported primarily by the National Science Foundation through the University of Minnesota MRSEC under Award Number DMR-2011401. FJB acknowledges the financial support of NSF: DMR-1928166 and the Carnegie corporation of New York. AK was supported by the NSF grant DMR-2037654.

	\appendix

	\section{Fitting procedure for transport}\label{app:fitting}

	In order to fit the transport data of the zigzag ribbons to the two-parameter scaling functions of the 1D random hopping chain, one could potentially use the distribution Eq.~\ref{eqn:gamma_dsbn}. One could perform maximum likelihood fits to obtain the parameters $\delta$ and $\gamma$. However, analytic expressions for these quantities in terms of the relevant scaling parameters $s$ and $r$ is only known when $r \ll 1$. Due to the suppression of backscattering in the zigzag ribbons, the majority of transport does not fall in this regime. To circumvent this issue, one can use the numerical interpolating functions for relevant transport quantities as a function of $s, r$ that are available for the 1D chain \footnote{The datasets can be made available on reasonable request}.

\begin{figure}
		\centering
		\includegraphics[width=\linewidth]{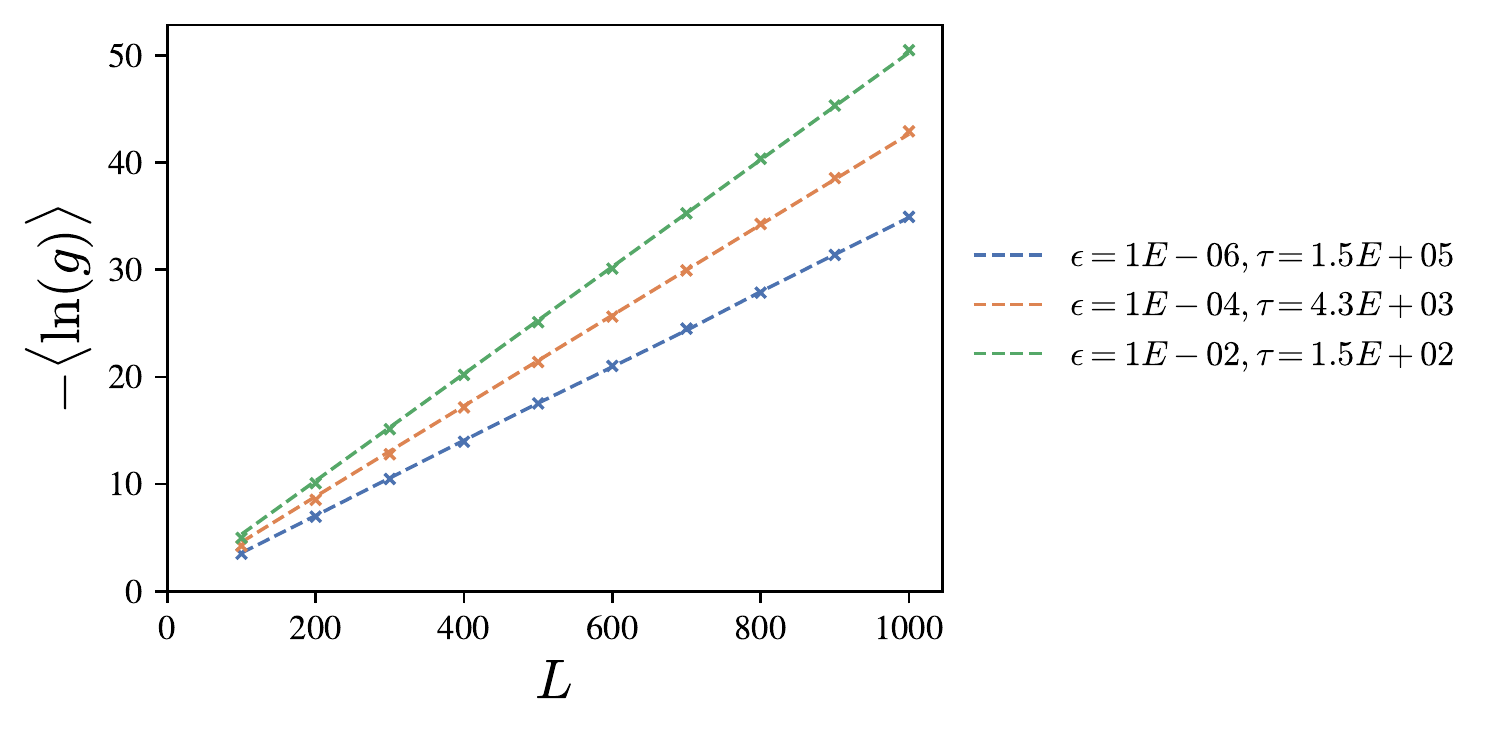}
		\caption{Fits of $\expval{\ln(g)}$ vs $L$ to the data obtained for the 1D chain are shown for a zigzag chain with $W=4$ and a few different energies. The crosses mark the numerical data and the dashed lines show the fits, with the relaxation time $\tau$ obtained from the fit shown in the legend.}
		\label{fig:zz-transport-fit}
	\end{figure}

	In this case $\expval{\ln(g)}$ is considered, as it has good self-averaging properties and is a well-behaved function. Here $\expval{\dots}$ dentoes averaging over disorder realizations. Given a zigzag ribbon of certain width $W$ and energy of incoming electrons $\epsilon$, one can fit $\expval{\ln(g)}$ vs $L$ to the data from the 1D chain. This allows one to extract $l$ and $\tau$. Note that there is only one free fitting parameter since $l = v_F \tau$, where the velocity $v_F$ is known at a given energy. Figure \ref{fig:zz-transport-fit} shows a few fits for $W=4$ as the energy is varied. The quality of the fits is generally good.

\bibliographystyle{apsrev4-2}
\bibliography{references_zigzag}

\begin{thebibliography}{57}%
\makeatletter
\providecommand \@ifxundefined [1]{%
 \@ifx{#1\undefined}
}%
\providecommand \@ifnum [1]{%
 \ifnum #1\expandafter \@firstoftwo
 \else \expandafter \@secondoftwo
 \fi
}%
\providecommand \@ifx [1]{%
 \ifx #1\expandafter \@firstoftwo
 \else \expandafter \@secondoftwo
 \fi
}%
\providecommand \natexlab [1]{#1}%
\providecommand \enquote  [1]{``#1''}%
\providecommand \bibnamefont  [1]{#1}%
\providecommand \bibfnamefont [1]{#1}%
\providecommand \citenamefont [1]{#1}%
\providecommand \href@noop [0]{\@secondoftwo}%
\providecommand \href [0]{\begingroup \@sanitize@url \@href}%
\providecommand \@href[1]{\@@startlink{#1}\@@href}%
\providecommand \@@href[1]{\endgroup#1\@@endlink}%
\providecommand \@sanitize@url [0]{\catcode `\\12\catcode `\$12\catcode
  `\&12\catcode `\#12\catcode `\^12\catcode `\_12\catcode `\%12\relax}%
\providecommand \@@startlink[1]{}%
\providecommand \@@endlink[0]{}%
\providecommand \url  [0]{\begingroup\@sanitize@url \@url }%
\providecommand \@url [1]{\endgroup\@href {#1}{\urlprefix }}%
\providecommand \urlprefix  [0]{URL }%
\providecommand \Eprint [0]{\href }%
\providecommand \doibase [0]{https://doi.org/}%
\providecommand \selectlanguage [0]{\@gobble}%
\providecommand \bibinfo  [0]{\@secondoftwo}%
\providecommand \bibfield  [0]{\@secondoftwo}%
\providecommand \translation [1]{[#1]}%
\providecommand \BibitemOpen [0]{}%
\providecommand \bibitemStop [0]{}%
\providecommand \bibitemNoStop [0]{.\EOS\space}%
\providecommand \EOS [0]{\spacefactor3000\relax}%
\providecommand \BibitemShut  [1]{\csname bibitem#1\endcsname}%
\let\auto@bib@innerbib\@empty
\bibitem [{\citenamefont {Anderson}(1958)}]{AndersonLocalization}%
  \BibitemOpen
  \bibfield  {author} {\bibinfo {author} {\bibfnamefont {P.~W.}\ \bibnamefont
  {Anderson}},\ }\href {https://doi.org/10.1103/PhysRev.109.1492} {\bibfield
  {journal} {\bibinfo  {journal} {Phys. Rev.}\ }\textbf {\bibinfo {volume}
  {109}},\ \bibinfo {pages} {1492} (\bibinfo {year} {1958})}\BibitemShut
  {NoStop}%
\bibitem [{\citenamefont {Lee}\ and\ \citenamefont
  {Ramakrishnan}(1985)}]{LeeRamakrishnanReview}%
  \BibitemOpen
  \bibfield  {author} {\bibinfo {author} {\bibfnamefont {P.~A.}\ \bibnamefont
  {Lee}}\ and\ \bibinfo {author} {\bibfnamefont {T.~V.}\ \bibnamefont
  {Ramakrishnan}},\ }\href {https://doi.org/10.1103/RevModPhys.57.287}
  {\bibfield  {journal} {\bibinfo  {journal} {Rev. Mod. Phys.}\ }\textbf
  {\bibinfo {volume} {57}},\ \bibinfo {pages} {287} (\bibinfo {year}
  {1985})}\BibitemShut {NoStop}%
\bibitem [{\citenamefont {Kramer}\ and\ \citenamefont
  {MacKinnon}(1993)}]{KramerMackinnonReview}%
  \BibitemOpen
  \bibfield  {author} {\bibinfo {author} {\bibfnamefont {B.}~\bibnamefont
  {Kramer}}\ and\ \bibinfo {author} {\bibfnamefont {A.}~\bibnamefont
  {MacKinnon}},\ }\href {https://doi.org/10.1088/0034-4885/56/12/001}
  {\bibfield  {journal} {\bibinfo  {journal} {Reports on Progress in Physics}\
  }\textbf {\bibinfo {volume} {56}},\ \bibinfo {pages} {1469} (\bibinfo {year}
  {1993})}\BibitemShut {NoStop}%
\bibitem [{\citenamefont {Evers}\ and\ \citenamefont
  {Mirlin}(2008)}]{EversMirlinReview}%
  \BibitemOpen
  \bibfield  {author} {\bibinfo {author} {\bibfnamefont {F.}~\bibnamefont
  {Evers}}\ and\ \bibinfo {author} {\bibfnamefont {A.~D.}\ \bibnamefont
  {Mirlin}},\ }\href {https://doi.org/10.1103/RevModPhys.80.1355} {\bibfield
  {journal} {\bibinfo  {journal} {Rev. Mod. Phys.}\ }\textbf {\bibinfo {volume}
  {80}},\ \bibinfo {pages} {1355} (\bibinfo {year} {2008})}\BibitemShut
  {NoStop}%
\bibitem [{\citenamefont {Abrahams}\ \emph {et~al.}(1979)\citenamefont
  {Abrahams}, \citenamefont {Anderson}, \citenamefont {Licciardello},\ and\
  \citenamefont {Ramakrishnan}}]{ScalingTheoryLoc}%
  \BibitemOpen
  \bibfield  {author} {\bibinfo {author} {\bibfnamefont {E.}~\bibnamefont
  {Abrahams}}, \bibinfo {author} {\bibfnamefont {P.~W.}\ \bibnamefont
  {Anderson}}, \bibinfo {author} {\bibfnamefont {D.~C.}\ \bibnamefont
  {Licciardello}},\ and\ \bibinfo {author} {\bibfnamefont {T.~V.}\ \bibnamefont
  {Ramakrishnan}},\ }\href {https://doi.org/10.1103/PhysRevLett.42.673}
  {\bibfield  {journal} {\bibinfo  {journal} {Phys. Rev. Lett.}\ }\textbf
  {\bibinfo {volume} {42}},\ \bibinfo {pages} {673} (\bibinfo {year}
  {1979})}\BibitemShut {NoStop}%
\bibitem [{\citenamefont {{Gor'kov}}\ \emph {et~al.}(1979)\citenamefont
  {{Gor'kov}}, \citenamefont {{Larkin}},\ and\ \citenamefont
  {{Khmel'Nitski{\v{i}}}}}]{GorkovLarkinScaling}%
  \BibitemOpen
  \bibfield  {author} {\bibinfo {author} {\bibfnamefont {L.~P.}\ \bibnamefont
  {{Gor'kov}}}, \bibinfo {author} {\bibfnamefont {A.~I.}\ \bibnamefont
  {{Larkin}}},\ and\ \bibinfo {author} {\bibfnamefont {D.~E.}\ \bibnamefont
  {{Khmel'Nitski{\v{i}}}}},\ }\href@noop {} {\bibfield  {journal} {\bibinfo
  {journal} {Soviet Journal of Experimental and Theoretical Physics Letters}\
  }\textbf {\bibinfo {volume} {30}},\ \bibinfo {pages} {228} (\bibinfo {year}
  {1979})}\BibitemShut {NoStop}%
\bibitem [{\citenamefont {Dyson}(1953)}]{DysonDOS}%
  \BibitemOpen
  \bibfield  {author} {\bibinfo {author} {\bibfnamefont {F.~J.}\ \bibnamefont
  {Dyson}},\ }\href {https://doi.org/10.1103/PhysRev.92.1331} {\bibfield
  {journal} {\bibinfo  {journal} {Phys. Rev.}\ }\textbf {\bibinfo {volume}
  {92}},\ \bibinfo {pages} {1331} (\bibinfo {year} {1953})}\BibitemShut
  {NoStop}%
\bibitem [{\citenamefont {Su}\ \emph {et~al.}(1979)\citenamefont {Su},
  \citenamefont {Schrieffer},\ and\ \citenamefont {Heeger}}]{SSH}%
  \BibitemOpen
  \bibfield  {author} {\bibinfo {author} {\bibfnamefont {W.~P.}\ \bibnamefont
  {Su}}, \bibinfo {author} {\bibfnamefont {J.~R.}\ \bibnamefont {Schrieffer}},\
  and\ \bibinfo {author} {\bibfnamefont {A.~J.}\ \bibnamefont {Heeger}},\
  }\href {https://doi.org/10.1103/PhysRevLett.42.1698} {\bibfield  {journal}
  {\bibinfo  {journal} {Phys. Rev. Lett.}\ }\textbf {\bibinfo {volume} {42}},\
  \bibinfo {pages} {1698} (\bibinfo {year} {1979})}\BibitemShut {NoStop}%
\bibitem [{\citenamefont {Zak}(1989)}]{ZakPhase}%
  \BibitemOpen
  \bibfield  {author} {\bibinfo {author} {\bibfnamefont {J.}~\bibnamefont
  {Zak}},\ }\href {https://doi.org/10.1103/PhysRevLett.62.2747} {\bibfield
  {journal} {\bibinfo  {journal} {Phys. Rev. Lett.}\ }\textbf {\bibinfo
  {volume} {62}},\ \bibinfo {pages} {2747} (\bibinfo {year}
  {1989})}\BibitemShut {NoStop}%
\bibitem [{\citenamefont {Motrunich}\ \emph {et~al.}(2001)\citenamefont
  {Motrunich}, \citenamefont {Damle},\ and\ \citenamefont
  {Huse}}]{MotrunichDamleHuse}%
  \BibitemOpen
  \bibfield  {author} {\bibinfo {author} {\bibfnamefont {O.}~\bibnamefont
  {Motrunich}}, \bibinfo {author} {\bibfnamefont {K.}~\bibnamefont {Damle}},\
  and\ \bibinfo {author} {\bibfnamefont {D.~A.}\ \bibnamefont {Huse}},\ }\href
  {https://doi.org/10.1103/PhysRevB.63.224204} {\bibfield  {journal} {\bibinfo
  {journal} {Phys. Rev. B}\ }\textbf {\bibinfo {volume} {63}},\ \bibinfo
  {pages} {224204} (\bibinfo {year} {2001})}\BibitemShut {NoStop}%
\bibitem [{\citenamefont {Balents}\ and\ \citenamefont
  {Fisher}(1997)}]{BalentsFisher}%
  \BibitemOpen
  \bibfield  {author} {\bibinfo {author} {\bibfnamefont {L.}~\bibnamefont
  {Balents}}\ and\ \bibinfo {author} {\bibfnamefont {M.~P.~A.}\ \bibnamefont
  {Fisher}},\ }\href {https://doi.org/10.1103/PhysRevB.56.12970} {\bibfield
  {journal} {\bibinfo  {journal} {Phys. Rev. B}\ }\textbf {\bibinfo {volume}
  {56}},\ \bibinfo {pages} {12970} (\bibinfo {year} {1997})}\BibitemShut
  {NoStop}%
\bibitem [{\citenamefont {Mondragon-Shem}\ \emph {et~al.}(2014)\citenamefont
  {Mondragon-Shem}, \citenamefont {Hughes}, \citenamefont {Song},\ and\
  \citenamefont {Prodan}}]{Prodan2014}%
  \BibitemOpen
  \bibfield  {author} {\bibinfo {author} {\bibfnamefont {I.}~\bibnamefont
  {Mondragon-Shem}}, \bibinfo {author} {\bibfnamefont {T.~L.}\ \bibnamefont
  {Hughes}}, \bibinfo {author} {\bibfnamefont {J.}~\bibnamefont {Song}},\ and\
  \bibinfo {author} {\bibfnamefont {E.}~\bibnamefont {Prodan}},\ }\href
  {https://doi.org/10.1103/PhysRevLett.113.046802} {\bibfield  {journal}
  {\bibinfo  {journal} {Phys. Rev. Lett.}\ }\textbf {\bibinfo {volume} {113}},\
  \bibinfo {pages} {046802} (\bibinfo {year} {2014})}\BibitemShut {NoStop}%
\bibitem [{\citenamefont {Bagrets}\ \emph {et~al.}(2016)\citenamefont
  {Bagrets}, \citenamefont {Altland},\ and\ \citenamefont
  {Kamenev}}]{Kamenev2016}%
  \BibitemOpen
  \bibfield  {author} {\bibinfo {author} {\bibfnamefont {D.}~\bibnamefont
  {Bagrets}}, \bibinfo {author} {\bibfnamefont {A.}~\bibnamefont {Altland}},\
  and\ \bibinfo {author} {\bibfnamefont {A.}~\bibnamefont {Kamenev}},\ }\href
  {https://doi.org/10.1103/PhysRevLett.117.196801} {\bibfield  {journal}
  {\bibinfo  {journal} {Phys. Rev. Lett.}\ }\textbf {\bibinfo {volume} {117}},\
  \bibinfo {pages} {196801} (\bibinfo {year} {2016})}\BibitemShut {NoStop}%
\bibitem [{\citenamefont {Shankar}\ and\ \citenamefont
  {Murthy}(1987)}]{ShankarXY}%
  \BibitemOpen
  \bibfield  {author} {\bibinfo {author} {\bibfnamefont {R.}~\bibnamefont
  {Shankar}}\ and\ \bibinfo {author} {\bibfnamefont {G.}~\bibnamefont
  {Murthy}},\ }\href {https://doi.org/10.1103/PhysRevB.36.536} {\bibfield
  {journal} {\bibinfo  {journal} {Phys. Rev. B}\ }\textbf {\bibinfo {volume}
  {36}},\ \bibinfo {pages} {536} (\bibinfo {year} {1987})}\BibitemShut
  {NoStop}%
\bibitem [{\citenamefont {McKenzie}(1996)}]{McKenzieQPT}%
  \BibitemOpen
  \bibfield  {author} {\bibinfo {author} {\bibfnamefont {R.~H.}\ \bibnamefont
  {McKenzie}},\ }\href {https://doi.org/10.1103/PhysRevLett.77.4804} {\bibfield
   {journal} {\bibinfo  {journal} {Phys. Rev. Lett.}\ }\textbf {\bibinfo
  {volume} {77}},\ \bibinfo {pages} {4804} (\bibinfo {year}
  {1996})}\BibitemShut {NoStop}%
\bibitem [{\citenamefont {N.}(1982)}]{Dorokhov82}%
  \BibitemOpen
  \bibfield  {author} {\bibinfo {author} {\bibfnamefont {D.~O.}\ \bibnamefont
  {N.}},\ }\href@noop {} {\bibfield  {journal} {\bibinfo  {journal} {Pis’ma
  Zh. Eksp. Teor. Fiz.}\ }\textbf {\bibinfo {volume} {36}},\ \bibinfo {pages}
  {259} (\bibinfo {year} {1982})},\ \bibinfo {note} {jETP Lett. 36,
  318}\BibitemShut {NoStop}%
\bibitem [{\citenamefont {Mello}\ \emph {et~al.}(1988)\citenamefont {Mello},
  \citenamefont {Pereyra},\ and\ \citenamefont {Kumar}}]{MPK}%
  \BibitemOpen
  \bibfield  {author} {\bibinfo {author} {\bibfnamefont {P.}~\bibnamefont
  {Mello}}, \bibinfo {author} {\bibfnamefont {P.}~\bibnamefont {Pereyra}},\
  and\ \bibinfo {author} {\bibfnamefont {N.}~\bibnamefont {Kumar}},\ }\href
  {https://doi.org/https://doi.org/10.1016/0003-4916(88)90169-8} {\bibfield
  {journal} {\bibinfo  {journal} {Annals of Physics}\ }\textbf {\bibinfo
  {volume} {181}},\ \bibinfo {pages} {290} (\bibinfo {year}
  {1988})}\BibitemShut {NoStop}%
\bibitem [{\citenamefont {Mace\^do}\ and\ \citenamefont
  {Chalker}(1992)}]{MacedoChalker}%
  \BibitemOpen
  \bibfield  {author} {\bibinfo {author} {\bibfnamefont {A.~M.~S.}\
  \bibnamefont {Mace\^do}}\ and\ \bibinfo {author} {\bibfnamefont {J.~T.}\
  \bibnamefont {Chalker}},\ }\href {https://doi.org/10.1103/PhysRevB.46.14985}
  {\bibfield  {journal} {\bibinfo  {journal} {Phys. Rev. B}\ }\textbf {\bibinfo
  {volume} {46}},\ \bibinfo {pages} {14985} (\bibinfo {year}
  {1992})}\BibitemShut {NoStop}%
\bibitem [{\citenamefont {Beenakker}(1997)}]{BeenakkerRMT}%
  \BibitemOpen
  \bibfield  {author} {\bibinfo {author} {\bibfnamefont {C.~W.~J.}\
  \bibnamefont {Beenakker}},\ }\href
  {https://doi.org/10.1103/RevModPhys.69.731} {\bibfield  {journal} {\bibinfo
  {journal} {Rev. Mod. Phys.}\ }\textbf {\bibinfo {volume} {69}},\ \bibinfo
  {pages} {731} (\bibinfo {year} {1997})}\BibitemShut {NoStop}%
\bibitem [{\citenamefont {Brouwer}\ \emph {et~al.}(1998)\citenamefont
  {Brouwer}, \citenamefont {Mudry}, \citenamefont {Simons},\ and\ \citenamefont
  {Altland}}]{Deloc1DChains}%
  \BibitemOpen
  \bibfield  {author} {\bibinfo {author} {\bibfnamefont {P.~W.}\ \bibnamefont
  {Brouwer}}, \bibinfo {author} {\bibfnamefont {C.}~\bibnamefont {Mudry}},
  \bibinfo {author} {\bibfnamefont {B.~D.}\ \bibnamefont {Simons}},\ and\
  \bibinfo {author} {\bibfnamefont {A.}~\bibnamefont {Altland}},\ }\href
  {https://doi.org/10.1103/PhysRevLett.81.862} {\bibfield  {journal} {\bibinfo
  {journal} {Phys. Rev. Lett.}\ }\textbf {\bibinfo {volume} {81}},\ \bibinfo
  {pages} {862} (\bibinfo {year} {1998})}\BibitemShut {NoStop}%
\bibitem [{\citenamefont {Mudry}\ \emph {et~al.}(1999)\citenamefont {Mudry},
  \citenamefont {Brouwer},\ and\ \citenamefont {Furusaki}}]{MudryRandomFlux}%
  \BibitemOpen
  \bibfield  {author} {\bibinfo {author} {\bibfnamefont {C.}~\bibnamefont
  {Mudry}}, \bibinfo {author} {\bibfnamefont {P.~W.}\ \bibnamefont {Brouwer}},\
  and\ \bibinfo {author} {\bibfnamefont {A.}~\bibnamefont {Furusaki}},\ }\href
  {https://doi.org/10.1103/PhysRevB.59.13221} {\bibfield  {journal} {\bibinfo
  {journal} {Phys. Rev. B}\ }\textbf {\bibinfo {volume} {59}},\ \bibinfo
  {pages} {13221} (\bibinfo {year} {1999})}\BibitemShut {NoStop}%
\bibitem [{\citenamefont {Altland}\ and\ \citenamefont
  {Zirnbauer}(1997)}]{AltlandZirnbauer}%
  \BibitemOpen
  \bibfield  {author} {\bibinfo {author} {\bibfnamefont {A.}~\bibnamefont
  {Altland}}\ and\ \bibinfo {author} {\bibfnamefont {M.~R.}\ \bibnamefont
  {Zirnbauer}},\ }\href {https://doi.org/10.1103/PhysRevB.55.1142} {\bibfield
  {journal} {\bibinfo  {journal} {Phys. Rev. B}\ }\textbf {\bibinfo {volume}
  {55}},\ \bibinfo {pages} {1142} (\bibinfo {year} {1997})}\BibitemShut
  {NoStop}%
\bibitem [{\citenamefont {Ryu}\ \emph {et~al.}(2010)\citenamefont {Ryu},
  \citenamefont {Schnyder}, \citenamefont {Furusaki},\ and\ \citenamefont
  {Ludwig}}]{RyuTenFoldWay}%
  \BibitemOpen
  \bibfield  {author} {\bibinfo {author} {\bibfnamefont {S.}~\bibnamefont
  {Ryu}}, \bibinfo {author} {\bibfnamefont {A.~P.}\ \bibnamefont {Schnyder}},
  \bibinfo {author} {\bibfnamefont {A.}~\bibnamefont {Furusaki}},\ and\
  \bibinfo {author} {\bibfnamefont {A.~W.~W.}\ \bibnamefont {Ludwig}},\ }\href
  {https://doi.org/10.1088/1367-2630/12/6/065010} {\bibfield  {journal}
  {\bibinfo  {journal} {New Journal of Physics}\ }\textbf {\bibinfo {volume}
  {12}},\ \bibinfo {pages} {065010} (\bibinfo {year} {2010})}\BibitemShut
  {NoStop}%
\bibitem [{\citenamefont {Kitaev}(2009)}]{KitaevClassification}%
  \BibitemOpen
  \bibfield  {author} {\bibinfo {author} {\bibfnamefont {A.}~\bibnamefont
  {Kitaev}},\ }\href {https://doi.org/10.1063/1.3149495} {\bibfield  {journal}
  {\bibinfo  {journal} {AIP Conference Proceedings}\ }\textbf {\bibinfo
  {volume} {1134}},\ \bibinfo {pages} {22} (\bibinfo {year}
  {2009})}\BibitemShut {NoStop}%
\bibitem [{\citenamefont {Ludwig}(2015)}]{LudwigClassification}%
  \BibitemOpen
  \bibfield  {author} {\bibinfo {author} {\bibfnamefont {A.~W.~W.}\
  \bibnamefont {Ludwig}},\ }\href
  {https://doi.org/10.1088/0031-8949/2015/t168/014001} {\bibfield  {journal}
  {\bibinfo  {journal} {Physica Scripta}\ }\textbf {\bibinfo {volume} {T168}},\
  \bibinfo {pages} {014001} (\bibinfo {year} {2015})}\BibitemShut {NoStop}%
\bibitem [{\citenamefont {{Brouwer}}\ \emph {et~al.}(2005)\citenamefont
  {{Brouwer}}, \citenamefont {{Furusaki}}, \citenamefont {{Mudry}},\ and\
  \citenamefont {{Ryu}}}]{BrouwerFPReview}%
  \BibitemOpen
  \bibfield  {author} {\bibinfo {author} {\bibfnamefont {P.~W.}\ \bibnamefont
  {{Brouwer}}}, \bibinfo {author} {\bibfnamefont {A.}~\bibnamefont
  {{Furusaki}}}, \bibinfo {author} {\bibfnamefont {C.}~\bibnamefont
  {{Mudry}}},\ and\ \bibinfo {author} {\bibfnamefont {S.}~\bibnamefont
  {{Ryu}}},\ }\href@noop {} {\bibfield  {journal} {\bibinfo  {journal} {arXiv
  e-prints}\ ,\ \bibinfo {eid} {cond-mat/0511622}} (\bibinfo {year} {2005})},\
  \Eprint {https://arxiv.org/abs/cond-mat/0511622} {arXiv:cond-mat/0511622
  [cond-mat.mes-hall]} \BibitemShut {NoStop}%
\bibitem [{\citenamefont {Brouwer}\ \emph {et~al.}(2000)\citenamefont
  {Brouwer}, \citenamefont {Furusaki}, \citenamefont {Gruzberg},\ and\
  \citenamefont {Mudry}}]{LocalizationDelocalizationDirtySC}%
  \BibitemOpen
  \bibfield  {author} {\bibinfo {author} {\bibfnamefont {P.~W.}\ \bibnamefont
  {Brouwer}}, \bibinfo {author} {\bibfnamefont {A.}~\bibnamefont {Furusaki}},
  \bibinfo {author} {\bibfnamefont {I.~A.}\ \bibnamefont {Gruzberg}},\ and\
  \bibinfo {author} {\bibfnamefont {C.}~\bibnamefont {Mudry}},\ }\href
  {https://doi.org/10.1103/PhysRevLett.85.1064} {\bibfield  {journal} {\bibinfo
   {journal} {Phys. Rev. Lett.}\ }\textbf {\bibinfo {volume} {85}},\ \bibinfo
  {pages} {1064} (\bibinfo {year} {2000})}\BibitemShut {NoStop}%
\bibitem [{\citenamefont {Gruzberg}\ \emph {et~al.}(2005)\citenamefont
  {Gruzberg}, \citenamefont {Read},\ and\ \citenamefont
  {Vishveshwara}}]{GruzbergSuperuniversality}%
  \BibitemOpen
  \bibfield  {author} {\bibinfo {author} {\bibfnamefont {I.~A.}\ \bibnamefont
  {Gruzberg}}, \bibinfo {author} {\bibfnamefont {N.}~\bibnamefont {Read}},\
  and\ \bibinfo {author} {\bibfnamefont {S.}~\bibnamefont {Vishveshwara}},\
  }\href {https://doi.org/10.1103/PhysRevB.71.245124} {\bibfield  {journal}
  {\bibinfo  {journal} {Phys. Rev. B}\ }\textbf {\bibinfo {volume} {71}},\
  \bibinfo {pages} {245124} (\bibinfo {year} {2005})}\BibitemShut {NoStop}%
\bibitem [{\citenamefont {Bernevig}\ and\ \citenamefont
  {Hughes}(2013)}]{BernevigHughesGrapheneChapter}%
  \BibitemOpen
  \bibfield  {author} {\bibinfo {author} {\bibfnamefont {B.~A.}\ \bibnamefont
  {Bernevig}}\ and\ \bibinfo {author} {\bibfnamefont {T.~L.}\ \bibnamefont
  {Hughes}},\ }\bibinfo {title} {Graphene},\ in\ \href
  {http://www.jstor.org/stable/j.ctt19cc2gc.9} {\emph {\bibinfo {booktitle}
  {Topological Insulators and Topological Superconductors}}}\ (\bibinfo
  {publisher} {Princeton University Press},\ \bibinfo {year} {2013})\ pp.\
  \bibinfo {pages} {70--90}\BibitemShut {NoStop}%
\bibitem [{\citenamefont {Wakabayashi}\ \emph {et~al.}(2010)\citenamefont
  {Wakabayashi}, \citenamefont {ichi Sasaki}, \citenamefont {Nakanishi},\ and\
  \citenamefont {Enoki}}]{WakabayashiGrapheneRibbons}%
  \BibitemOpen
  \bibfield  {author} {\bibinfo {author} {\bibfnamefont {K.}~\bibnamefont
  {Wakabayashi}}, \bibinfo {author} {\bibfnamefont {K.}~\bibnamefont {ichi
  Sasaki}}, \bibinfo {author} {\bibfnamefont {T.}~\bibnamefont {Nakanishi}},\
  and\ \bibinfo {author} {\bibfnamefont {T.}~\bibnamefont {Enoki}},\ }\href
  {https://doi.org/10.1088/1468-6996/11/5/054504} {\bibfield  {journal}
  {\bibinfo  {journal} {Science and Technology of Advanced Materials}\ }\textbf
  {\bibinfo {volume} {11}},\ \bibinfo {pages} {054504} (\bibinfo {year}
  {2010})},\ \bibinfo {note} {pMID: 27877361}\BibitemShut {NoStop}%
\bibitem [{\citenamefont {Titov}\ \emph {et~al.}(2001)\citenamefont {Titov},
  \citenamefont {Brouwer}, \citenamefont {Furusaki},\ and\ \citenamefont
  {Mudry}}]{TitovDOS}%
  \BibitemOpen
  \bibfield  {author} {\bibinfo {author} {\bibfnamefont {M.}~\bibnamefont
  {Titov}}, \bibinfo {author} {\bibfnamefont {P.~W.}\ \bibnamefont {Brouwer}},
  \bibinfo {author} {\bibfnamefont {A.}~\bibnamefont {Furusaki}},\ and\
  \bibinfo {author} {\bibfnamefont {C.}~\bibnamefont {Mudry}},\ }\href
  {https://doi.org/10.1103/PhysRevB.63.235318} {\bibfield  {journal} {\bibinfo
  {journal} {Phys. Rev. B}\ }\textbf {\bibinfo {volume} {63}},\ \bibinfo
  {pages} {235318} (\bibinfo {year} {2001})}\BibitemShut {NoStop}%
\bibitem [{\citenamefont {Theodorou}\ and\ \citenamefont
  {Cohen}(1976)}]{Theodorou1DChain}%
  \BibitemOpen
  \bibfield  {author} {\bibinfo {author} {\bibfnamefont {G.}~\bibnamefont
  {Theodorou}}\ and\ \bibinfo {author} {\bibfnamefont {M.~H.}\ \bibnamefont
  {Cohen}},\ }\href {https://doi.org/10.1103/PhysRevB.13.4597} {\bibfield
  {journal} {\bibinfo  {journal} {Phys. Rev. B}\ }\textbf {\bibinfo {volume}
  {13}},\ \bibinfo {pages} {4597} (\bibinfo {year} {1976})}\BibitemShut
  {NoStop}%
\bibitem [{\citenamefont {Eggarter}\ and\ \citenamefont
  {Riedinger}(1978)}]{EggarterLocDivergence}%
  \BibitemOpen
  \bibfield  {author} {\bibinfo {author} {\bibfnamefont {T.~P.}\ \bibnamefont
  {Eggarter}}\ and\ \bibinfo {author} {\bibfnamefont {R.}~\bibnamefont
  {Riedinger}},\ }\href {https://doi.org/10.1103/PhysRevB.18.569} {\bibfield
  {journal} {\bibinfo  {journal} {Phys. Rev. B}\ }\textbf {\bibinfo {volume}
  {18}},\ \bibinfo {pages} {569} (\bibinfo {year} {1978})}\BibitemShut
  {NoStop}%
\bibitem [{\citenamefont {Ziman}(1982)}]{ZimanLocDivergence}%
  \BibitemOpen
  \bibfield  {author} {\bibinfo {author} {\bibfnamefont {T.~A.~L.}\
  \bibnamefont {Ziman}},\ }\href {https://doi.org/10.1103/PhysRevLett.49.337}
  {\bibfield  {journal} {\bibinfo  {journal} {Phys. Rev. Lett.}\ }\textbf
  {\bibinfo {volume} {49}},\ \bibinfo {pages} {337} (\bibinfo {year}
  {1982})}\BibitemShut {NoStop}%
\bibitem [{\citenamefont {Fisher}(1995)}]{DFisherIsingChains}%
  \BibitemOpen
  \bibfield  {author} {\bibinfo {author} {\bibfnamefont {D.~S.}\ \bibnamefont
  {Fisher}},\ }\href {https://doi.org/10.1103/PhysRevB.51.6411} {\bibfield
  {journal} {\bibinfo  {journal} {Phys. Rev. B}\ }\textbf {\bibinfo {volume}
  {51}},\ \bibinfo {pages} {6411} (\bibinfo {year} {1995})}\BibitemShut
  {NoStop}%
\bibitem [{\citenamefont {Altland}\ \emph {et~al.}(2014)\citenamefont
  {Altland}, \citenamefont {Bagrets}, \citenamefont {Fritz}, \citenamefont
  {Kamenev},\ and\ \citenamefont {Schmiedt}}]{Kamenev2014}%
  \BibitemOpen
  \bibfield  {author} {\bibinfo {author} {\bibfnamefont {A.}~\bibnamefont
  {Altland}}, \bibinfo {author} {\bibfnamefont {D.}~\bibnamefont {Bagrets}},
  \bibinfo {author} {\bibfnamefont {L.}~\bibnamefont {Fritz}}, \bibinfo
  {author} {\bibfnamefont {A.}~\bibnamefont {Kamenev}},\ and\ \bibinfo {author}
  {\bibfnamefont {H.}~\bibnamefont {Schmiedt}},\ }\href
  {https://doi.org/10.1103/PhysRevLett.112.206602} {\bibfield  {journal}
  {\bibinfo  {journal} {Phys. Rev. Lett.}\ }\textbf {\bibinfo {volume} {112}},\
  \bibinfo {pages} {206602} (\bibinfo {year} {2014})}\BibitemShut {NoStop}%
\bibitem [{\citenamefont {Ryu}\ \emph {et~al.}(2004)\citenamefont {Ryu},
  \citenamefont {Mudry},\ and\ \citenamefont {Furusaki}}]{RyuCrossover}%
  \BibitemOpen
  \bibfield  {author} {\bibinfo {author} {\bibfnamefont {S.}~\bibnamefont
  {Ryu}}, \bibinfo {author} {\bibfnamefont {C.}~\bibnamefont {Mudry}},\ and\
  \bibinfo {author} {\bibfnamefont {A.}~\bibnamefont {Furusaki}},\ }\href
  {https://doi.org/10.1103/PhysRevB.70.195329} {\bibfield  {journal} {\bibinfo
  {journal} {Phys. Rev. B}\ }\textbf {\bibinfo {volume} {70}},\ \bibinfo
  {pages} {195329} (\bibinfo {year} {2004})}\BibitemShut {NoStop}%
\bibitem [{\citenamefont {Kasturirangan}\ \emph {et~al.}(2022)\citenamefont
  {Kasturirangan}, \citenamefont {Kamenev},\ and\ \citenamefont
  {Burnell}}]{TwoParamScaling}%
  \BibitemOpen
  \bibfield  {author} {\bibinfo {author} {\bibfnamefont {S.}~\bibnamefont
  {Kasturirangan}}, \bibinfo {author} {\bibfnamefont {A.}~\bibnamefont
  {Kamenev}},\ and\ \bibinfo {author} {\bibfnamefont {F.~J.}\ \bibnamefont
  {Burnell}},\ }\href {https://doi.org/10.1103/PhysRevB.105.174204} {\bibfield
  {journal} {\bibinfo  {journal} {Phys. Rev. B}\ }\textbf {\bibinfo {volume}
  {105}},\ \bibinfo {pages} {174204} (\bibinfo {year} {2022})}\BibitemShut
  {NoStop}%
\bibitem [{\citenamefont {Wakabayashi}\ \emph {et~al.}(2007)\citenamefont
  {Wakabayashi}, \citenamefont {Takane},\ and\ \citenamefont
  {Sigrist}}]{WakabayashiPerfectlyConductingChannel}%
  \BibitemOpen
  \bibfield  {author} {\bibinfo {author} {\bibfnamefont {K.}~\bibnamefont
  {Wakabayashi}}, \bibinfo {author} {\bibfnamefont {Y.}~\bibnamefont
  {Takane}},\ and\ \bibinfo {author} {\bibfnamefont {M.}~\bibnamefont
  {Sigrist}},\ }\href {https://doi.org/10.1103/PhysRevLett.99.036601}
  {\bibfield  {journal} {\bibinfo  {journal} {Phys. Rev. Lett.}\ }\textbf
  {\bibinfo {volume} {99}},\ \bibinfo {pages} {036601} (\bibinfo {year}
  {2007})}\BibitemShut {NoStop}%
\bibitem [{\citenamefont {Wakabayashi}\ \emph {et~al.}(2009)\citenamefont
  {Wakabayashi}, \citenamefont {Takane}, \citenamefont {Yamamoto},\ and\
  \citenamefont {Sigrist}}]{WakabayashiPerfectlyConductingChannel2}%
  \BibitemOpen
  \bibfield  {author} {\bibinfo {author} {\bibfnamefont {K.}~\bibnamefont
  {Wakabayashi}}, \bibinfo {author} {\bibfnamefont {Y.}~\bibnamefont {Takane}},
  \bibinfo {author} {\bibfnamefont {M.}~\bibnamefont {Yamamoto}},\ and\
  \bibinfo {author} {\bibfnamefont {M.}~\bibnamefont {Sigrist}},\ }\href
  {https://doi.org/https://doi.org/10.1016/j.carbon.2008.09.040} {\bibfield
  {journal} {\bibinfo  {journal} {Carbon}\ }\textbf {\bibinfo {volume} {47}},\
  \bibinfo {pages} {124} (\bibinfo {year} {2009})}\BibitemShut {NoStop}%
\bibitem [{\citenamefont {Lima}\ \emph {et~al.}(2012)\citenamefont {Lima},
  \citenamefont {Pinheiro}, \citenamefont {Capaz}, \citenamefont {Lewenkopf},\
  and\ \citenamefont {Mucciolo}}]{MuccioloPerfectlyConductingChannel}%
  \BibitemOpen
  \bibfield  {author} {\bibinfo {author} {\bibfnamefont {L.~R.~F.}\
  \bibnamefont {Lima}}, \bibinfo {author} {\bibfnamefont {F.~A.}\ \bibnamefont
  {Pinheiro}}, \bibinfo {author} {\bibfnamefont {R.~B.}\ \bibnamefont {Capaz}},
  \bibinfo {author} {\bibfnamefont {C.~H.}\ \bibnamefont {Lewenkopf}},\ and\
  \bibinfo {author} {\bibfnamefont {E.~R.}\ \bibnamefont {Mucciolo}},\ }\href
  {https://doi.org/10.1103/PhysRevB.86.205111} {\bibfield  {journal} {\bibinfo
  {journal} {Phys. Rev. B}\ }\textbf {\bibinfo {volume} {86}},\ \bibinfo
  {pages} {205111} (\bibinfo {year} {2012})}\BibitemShut {NoStop}%
\bibitem [{\citenamefont {Lherbier}\ \emph {et~al.}(2008)\citenamefont
  {Lherbier}, \citenamefont {Biel}, \citenamefont {Niquet},\ and\ \citenamefont
  {Roche}}]{LocalizationinGrapheneLengthScales}%
  \BibitemOpen
  \bibfield  {author} {\bibinfo {author} {\bibfnamefont {A.}~\bibnamefont
  {Lherbier}}, \bibinfo {author} {\bibfnamefont {B.}~\bibnamefont {Biel}},
  \bibinfo {author} {\bibfnamefont {Y.-M.}\ \bibnamefont {Niquet}},\ and\
  \bibinfo {author} {\bibfnamefont {S.}~\bibnamefont {Roche}},\ }\href
  {https://doi.org/10.1103/PhysRevLett.100.036803} {\bibfield  {journal}
  {\bibinfo  {journal} {Phys. Rev. Lett.}\ }\textbf {\bibinfo {volume} {100}},\
  \bibinfo {pages} {036803} (\bibinfo {year} {2008})}\BibitemShut {NoStop}%
\bibitem [{\citenamefont {Schubert}\ and\ \citenamefont
  {Fehske}(2012)}]{MetalInsulatorTransitionGNR}%
  \BibitemOpen
  \bibfield  {author} {\bibinfo {author} {\bibfnamefont {G.}~\bibnamefont
  {Schubert}}\ and\ \bibinfo {author} {\bibfnamefont {H.}~\bibnamefont
  {Fehske}},\ }\href {https://doi.org/10.1103/PhysRevLett.108.066402}
  {\bibfield  {journal} {\bibinfo  {journal} {Phys. Rev. Lett.}\ }\textbf
  {\bibinfo {volume} {108}},\ \bibinfo {pages} {066402} (\bibinfo {year}
  {2012})}\BibitemShut {NoStop}%
\bibitem [{\citenamefont {Zhang}\ \emph {et~al.}(2009)\citenamefont {Zhang},
  \citenamefont {Hu}, \citenamefont {Bernevig}, \citenamefont {Wang},
  \citenamefont {Xie},\ and\ \citenamefont {Liu}}]{LocalizationandBKTGraphene}%
  \BibitemOpen
  \bibfield  {author} {\bibinfo {author} {\bibfnamefont {Y.-Y.}\ \bibnamefont
  {Zhang}}, \bibinfo {author} {\bibfnamefont {J.}~\bibnamefont {Hu}}, \bibinfo
  {author} {\bibfnamefont {B.~A.}\ \bibnamefont {Bernevig}}, \bibinfo {author}
  {\bibfnamefont {X.~R.}\ \bibnamefont {Wang}}, \bibinfo {author}
  {\bibfnamefont {X.~C.}\ \bibnamefont {Xie}},\ and\ \bibinfo {author}
  {\bibfnamefont {W.~M.}\ \bibnamefont {Liu}},\ }\href
  {https://doi.org/10.1103/PhysRevLett.102.106401} {\bibfield  {journal}
  {\bibinfo  {journal} {Phys. Rev. Lett.}\ }\textbf {\bibinfo {volume} {102}},\
  \bibinfo {pages} {106401} (\bibinfo {year} {2009})}\BibitemShut {NoStop}%
\bibitem [{\citenamefont {Mucciolo}\ \emph {et~al.}(2009)\citenamefont
  {Mucciolo}, \citenamefont {Castro~Neto},\ and\ \citenamefont
  {Lewenkopf}}]{MuccioloEdgeDisorder}%
  \BibitemOpen
  \bibfield  {author} {\bibinfo {author} {\bibfnamefont {E.~R.}\ \bibnamefont
  {Mucciolo}}, \bibinfo {author} {\bibfnamefont {A.~H.}\ \bibnamefont
  {Castro~Neto}},\ and\ \bibinfo {author} {\bibfnamefont {C.~H.}\ \bibnamefont
  {Lewenkopf}},\ }\href {https://doi.org/10.1103/PhysRevB.79.075407} {\bibfield
   {journal} {\bibinfo  {journal} {Phys. Rev. B}\ }\textbf {\bibinfo {volume}
  {79}},\ \bibinfo {pages} {075407} (\bibinfo {year} {2009})}\BibitemShut
  {NoStop}%
\bibitem [{\citenamefont {Dugaev}\ and\ \citenamefont
  {Katsnelson}(2013)}]{KatsnelsonEdgeDisorder}%
  \BibitemOpen
  \bibfield  {author} {\bibinfo {author} {\bibfnamefont {V.~K.}\ \bibnamefont
  {Dugaev}}\ and\ \bibinfo {author} {\bibfnamefont {M.~I.}\ \bibnamefont
  {Katsnelson}},\ }\href {https://doi.org/10.1103/PhysRevB.88.235432}
  {\bibfield  {journal} {\bibinfo  {journal} {Phys. Rev. B}\ }\textbf {\bibinfo
  {volume} {88}},\ \bibinfo {pages} {235432} (\bibinfo {year}
  {2013})}\BibitemShut {NoStop}%
\bibitem [{\citenamefont {Wimmer}\ \emph {et~al.}(2010)\citenamefont {Wimmer},
  \citenamefont {Akhmerov},\ and\ \citenamefont
  {Guinea}}]{GrapheneQuantumDotDisorder}%
  \BibitemOpen
  \bibfield  {author} {\bibinfo {author} {\bibfnamefont {M.}~\bibnamefont
  {Wimmer}}, \bibinfo {author} {\bibfnamefont {A.~R.}\ \bibnamefont
  {Akhmerov}},\ and\ \bibinfo {author} {\bibfnamefont {F.}~\bibnamefont
  {Guinea}},\ }\href {https://doi.org/10.1103/PhysRevB.82.045409} {\bibfield
  {journal} {\bibinfo  {journal} {Phys. Rev. B}\ }\textbf {\bibinfo {volume}
  {82}},\ \bibinfo {pages} {045409} (\bibinfo {year} {2010})}\BibitemShut
  {NoStop}%
\bibitem [{\citenamefont {Matsuura}\ \emph {et~al.}(2013)\citenamefont
  {Matsuura}, \citenamefont {Chang}, \citenamefont {Schnyder},\ and\
  \citenamefont {Ryu}}]{MatsuuraGaplessTopology}%
  \BibitemOpen
  \bibfield  {author} {\bibinfo {author} {\bibfnamefont {S.}~\bibnamefont
  {Matsuura}}, \bibinfo {author} {\bibfnamefont {P.-Y.}\ \bibnamefont {Chang}},
  \bibinfo {author} {\bibfnamefont {A.~P.}\ \bibnamefont {Schnyder}},\ and\
  \bibinfo {author} {\bibfnamefont {S.}~\bibnamefont {Ryu}},\ }\href
  {https://doi.org/10.1088/1367-2630/15/6/065001} {\bibfield  {journal}
  {\bibinfo  {journal} {New Journal of Physics}\ }\textbf {\bibinfo {volume}
  {15}},\ \bibinfo {pages} {065001} (\bibinfo {year} {2013})}\BibitemShut
  {NoStop}%
\bibitem [{\citenamefont {Wan}\ \emph {et~al.}(2011)\citenamefont {Wan},
  \citenamefont {Turner}, \citenamefont {Vishwanath},\ and\ \citenamefont
  {Savrasov}}]{WeylFermiArc}%
  \BibitemOpen
  \bibfield  {author} {\bibinfo {author} {\bibfnamefont {X.}~\bibnamefont
  {Wan}}, \bibinfo {author} {\bibfnamefont {A.~M.}\ \bibnamefont {Turner}},
  \bibinfo {author} {\bibfnamefont {A.}~\bibnamefont {Vishwanath}},\ and\
  \bibinfo {author} {\bibfnamefont {S.~Y.}\ \bibnamefont {Savrasov}},\ }\href
  {https://doi.org/10.1103/PhysRevB.83.205101} {\bibfield  {journal} {\bibinfo
  {journal} {Phys. Rev. B}\ }\textbf {\bibinfo {volume} {83}},\ \bibinfo
  {pages} {205101} (\bibinfo {year} {2011})}\BibitemShut {NoStop}%
\bibitem [{\citenamefont {Lv}\ \emph {et~al.}(2015)\citenamefont {Lv},
  \citenamefont {Weng}, \citenamefont {Fu}, \citenamefont {Wang}, \citenamefont
  {Miao}, \citenamefont {Ma}, \citenamefont {Richard}, \citenamefont {Huang},
  \citenamefont {Zhao}, \citenamefont {Chen}, \citenamefont {Fang},
  \citenamefont {Dai}, \citenamefont {Qian},\ and\ \citenamefont
  {Ding}}]{WeylExperiment1}%
  \BibitemOpen
  \bibfield  {author} {\bibinfo {author} {\bibfnamefont {B.~Q.}\ \bibnamefont
  {Lv}}, \bibinfo {author} {\bibfnamefont {H.~M.}\ \bibnamefont {Weng}},
  \bibinfo {author} {\bibfnamefont {B.~B.}\ \bibnamefont {Fu}}, \bibinfo
  {author} {\bibfnamefont {X.~P.}\ \bibnamefont {Wang}}, \bibinfo {author}
  {\bibfnamefont {H.}~\bibnamefont {Miao}}, \bibinfo {author} {\bibfnamefont
  {J.}~\bibnamefont {Ma}}, \bibinfo {author} {\bibfnamefont {P.}~\bibnamefont
  {Richard}}, \bibinfo {author} {\bibfnamefont {X.~C.}\ \bibnamefont {Huang}},
  \bibinfo {author} {\bibfnamefont {L.~X.}\ \bibnamefont {Zhao}}, \bibinfo
  {author} {\bibfnamefont {G.~F.}\ \bibnamefont {Chen}}, \bibinfo {author}
  {\bibfnamefont {Z.}~\bibnamefont {Fang}}, \bibinfo {author} {\bibfnamefont
  {X.}~\bibnamefont {Dai}}, \bibinfo {author} {\bibfnamefont {T.}~\bibnamefont
  {Qian}},\ and\ \bibinfo {author} {\bibfnamefont {H.}~\bibnamefont {Ding}},\
  }\href {https://doi.org/10.1103/PhysRevX.5.031013} {\bibfield  {journal}
  {\bibinfo  {journal} {Phys. Rev. X}\ }\textbf {\bibinfo {volume} {5}},\
  \bibinfo {pages} {031013} (\bibinfo {year} {2015})}\BibitemShut {NoStop}%
\bibitem [{\citenamefont {Xu}\ \emph {et~al.}(2015)\citenamefont {Xu},
  \citenamefont {Belopolski}, \citenamefont {Alidoust}, \citenamefont
  {Neupane}, \citenamefont {Bian}, \citenamefont {Zhang}, \citenamefont
  {Sankar}, \citenamefont {Chang}, \citenamefont {Yuan}, \citenamefont {Lee},
  \citenamefont {Huang}, \citenamefont {Zheng}, \citenamefont {Ma},
  \citenamefont {Sanchez}, \citenamefont {Wang}, \citenamefont {Bansil},
  \citenamefont {Chou}, \citenamefont {Shibayev}, \citenamefont {Lin},
  \citenamefont {Jia},\ and\ \citenamefont {Hasan}}]{WeylExperiment2}%
  \BibitemOpen
  \bibfield  {author} {\bibinfo {author} {\bibfnamefont {S.-Y.}\ \bibnamefont
  {Xu}}, \bibinfo {author} {\bibfnamefont {I.}~\bibnamefont {Belopolski}},
  \bibinfo {author} {\bibfnamefont {N.}~\bibnamefont {Alidoust}}, \bibinfo
  {author} {\bibfnamefont {M.}~\bibnamefont {Neupane}}, \bibinfo {author}
  {\bibfnamefont {G.}~\bibnamefont {Bian}}, \bibinfo {author} {\bibfnamefont
  {C.}~\bibnamefont {Zhang}}, \bibinfo {author} {\bibfnamefont
  {R.}~\bibnamefont {Sankar}}, \bibinfo {author} {\bibfnamefont
  {G.}~\bibnamefont {Chang}}, \bibinfo {author} {\bibfnamefont
  {Z.}~\bibnamefont {Yuan}}, \bibinfo {author} {\bibfnamefont {C.-C.}\
  \bibnamefont {Lee}}, \bibinfo {author} {\bibfnamefont {S.-M.}\ \bibnamefont
  {Huang}}, \bibinfo {author} {\bibfnamefont {H.}~\bibnamefont {Zheng}},
  \bibinfo {author} {\bibfnamefont {J.}~\bibnamefont {Ma}}, \bibinfo {author}
  {\bibfnamefont {D.~S.}\ \bibnamefont {Sanchez}}, \bibinfo {author}
  {\bibfnamefont {B.}~\bibnamefont {Wang}}, \bibinfo {author} {\bibfnamefont
  {A.}~\bibnamefont {Bansil}}, \bibinfo {author} {\bibfnamefont
  {F.}~\bibnamefont {Chou}}, \bibinfo {author} {\bibfnamefont {P.~P.}\
  \bibnamefont {Shibayev}}, \bibinfo {author} {\bibfnamefont {H.}~\bibnamefont
  {Lin}}, \bibinfo {author} {\bibfnamefont {S.}~\bibnamefont {Jia}},\ and\
  \bibinfo {author} {\bibfnamefont {M.~Z.}\ \bibnamefont {Hasan}},\ }\href
  {https://doi.org/10.1126/science.aaa9297} {\bibfield  {journal} {\bibinfo
  {journal} {Science}\ }\textbf {\bibinfo {volume} {349}},\ \bibinfo {pages}
  {613} (\bibinfo {year} {2015})}\BibitemShut {NoStop}%
\bibitem [{\citenamefont {Brouwer}\ \emph {et~al.}(2003)\citenamefont
  {Brouwer}, \citenamefont {Furusaki},\ and\ \citenamefont
  {Mudry}}]{BrouwerDirtySC}%
  \BibitemOpen
  \bibfield  {author} {\bibinfo {author} {\bibfnamefont {P.~W.}\ \bibnamefont
  {Brouwer}}, \bibinfo {author} {\bibfnamefont {A.}~\bibnamefont {Furusaki}},\
  and\ \bibinfo {author} {\bibfnamefont {C.}~\bibnamefont {Mudry}},\ }\href
  {https://doi.org/10.1103/PhysRevB.67.014530} {\bibfield  {journal} {\bibinfo
  {journal} {Phys. Rev. B}\ }\textbf {\bibinfo {volume} {67}},\ \bibinfo
  {pages} {014530} (\bibinfo {year} {2003})}\BibitemShut {NoStop}%
\bibitem [{\citenamefont {Groth}\ \emph {et~al.}(2014)\citenamefont {Groth},
  \citenamefont {Wimmer}, \citenamefont {Akhmerov},\ and\ \citenamefont
  {Waintal}}]{KWANT}%
  \BibitemOpen
  \bibfield  {author} {\bibinfo {author} {\bibfnamefont {C.~W.}\ \bibnamefont
  {Groth}}, \bibinfo {author} {\bibfnamefont {M.}~\bibnamefont {Wimmer}},
  \bibinfo {author} {\bibfnamefont {A.~R.}\ \bibnamefont {Akhmerov}},\ and\
  \bibinfo {author} {\bibfnamefont {X.}~\bibnamefont {Waintal}},\ }\href
  {https://doi.org/10.1088/1367-2630/16/6/063065} {\bibfield  {journal}
  {\bibinfo  {journal} {New Journal of Physics}\ }\textbf {\bibinfo {volume}
  {16}},\ \bibinfo {pages} {063065} (\bibinfo {year} {2014})}\BibitemShut
  {NoStop}%
\bibitem [{\citenamefont {Amestoy}\ \emph {et~al.}(2001)\citenamefont
  {Amestoy}, \citenamefont {Duff}, \citenamefont {Koster},\ and\ \citenamefont
  {L'Excellent}}]{MUMPS}%
  \BibitemOpen
  \bibfield  {author} {\bibinfo {author} {\bibfnamefont {P.}~\bibnamefont
  {Amestoy}}, \bibinfo {author} {\bibfnamefont {I.~S.}\ \bibnamefont {Duff}},
  \bibinfo {author} {\bibfnamefont {J.}~\bibnamefont {Koster}},\ and\ \bibinfo
  {author} {\bibfnamefont {J.-Y.}\ \bibnamefont {L'Excellent}},\ }\href@noop {}
  {\bibfield  {journal} {\bibinfo  {journal} {SIAM Journal on Matrix Analysis
  and Applications}\ }\textbf {\bibinfo {volume} {23}},\ \bibinfo {pages} {15}
  (\bibinfo {year} {2001})}\BibitemShut {NoStop}%
\bibitem [{\citenamefont {Abrikosov}(1981)}]{AbrikisovSolution}%
  \BibitemOpen
  \bibfield  {author} {\bibinfo {author} {\bibfnamefont {A.}~\bibnamefont
  {Abrikosov}},\ }\href
  {https://doi.org/https://doi.org/10.1016/0038-1098(81)91203-5} {\bibfield
  {journal} {\bibinfo  {journal} {Solid State Communications}\ }\textbf
  {\bibinfo {volume} {37}},\ \bibinfo {pages} {997} (\bibinfo {year}
  {1981})}\BibitemShut {NoStop}%
\bibitem [{\citenamefont {MacKinnon}(1985)}]{MacKinnonRGF}%
  \BibitemOpen
  \bibfield  {author} {\bibinfo {author} {\bibfnamefont {A.}~\bibnamefont
  {MacKinnon}},\ }\href {https://doi.org/10.1007/BF01328846} {\bibfield
  {journal} {\bibinfo  {journal} {Zeitschrift f{\"u}r Physik B Condensed
  Matter}\ }\textbf {\bibinfo {volume} {59}},\ \bibinfo {pages} {385} (\bibinfo
  {year} {1985})}\BibitemShut {NoStop}%
\bibitem [{Note1()}]{Note1}%
  \BibitemOpen
  \bibinfo {note} {The datasets can be made available on reasonable
  request}\BibitemShut {NoStop}%
\end{thebibliography}%

\end{document}